# *Symmetry-Based Classification of Chern Phases in Honeycomb Photonic Crystals*

*Rodrigo P. Câmara*[(1)], *Tatiana G. Rappoport*[(2,3)], *Mário G. Silveirinha*[(1)][*]

[(1)] *University of Lisbon – Instituto Superior Técnico and Instituto de Telecomunicações, Avenida Rovisco Pais, 1, 1049-001 Lisboa, Portugal,* rodrigo.camara@tecnico.ulisboa.pt

[(2)] *Centro de Física das Universidades do Minho e do Porto and Departamento de Física, Universidade do Minho, 4710-057 Braga, Portugal*

[(3)] *Instituto de Física, Universidade Federal do Rio de Janeiro, C.P. 68528, 21941-972 Rio de Janeiro, Rio de Janeiro, Brazil*

**Abstract**

In this work, we develop a symmetry-based classification of Chern phases in honeycomb photonic crystals, considering arbitrary nonreciprocal couplings compatible with energy conservation. Our analysis focuses on crystals formed through nonreciprocal perturbations of photonic graphene. These perturbations, which can have arbitrary spatial variations, are generally described by scalar and vector fields. Using a tight-binding model, we consider the most general nonreciprocal interactions, including gyromagnetic, pseudo-Tellegen, and moving medium responses, and examine how the corresponding nonreciprocal fields influence the crystal's topology. Our findings reveal that nonreciprocal interactions alone are insufficient to induce a topologically nontrivial phase. Instead, a nontrivial p6m component in the nonreciprocal fields is required to open a bandgap and achieve a non-zero Chern number. These results provide a symmetry-based roadmap for engineering photonic topological phases via nonreciprocal perturbations of photonic graphene, offering practical guidelines for designing topological phases in graphene-like photonic crystals.

---

[*] To whom correspondence should be addressed: E-mail: mario.silveirinha@tecnico.ulisboa.pt

# I. Introduction

Topology has captured the attention of physicists since the discovery of the quantum Hall effect [1] and its connection to the Chern (TKKN) number [2, 3]. It applies rather naturally in solids because the underlying first Brillouin Zone (FBZ) is a compact manifold without a boundary. The definitions of most topological numbers are built on this requirement [4-6], even though it may sometimes be relaxed [7-10]. As a result, the ideas of topology first developed in electronic systems have permeated other fields of physics also involving crystalline structures, such as mechanics [11], acoustics [12], and most notably photonics [13-15].

Photonic crystals have proven to be versatile platforms to create a plethora of topological phases, from Hall [16-19], spin-Hall [20-24], and higher-order [25-30] insulators to Dirac and Weyl semimetals [31-34]. In "light insulators", the topology of the bulk manifests itself at the surface in the form of unidirectional waveguiding of photons robust against deformations, fabrication defects, and other imperfections. Such reflectionless light channels are of great interest for boosting the transmission efficiency in light-based technologies.

The topological phases of photonic insulators are closely linked to symmetry. Several symmetry-based classifications of topological phases have been put forward [35, 36]. The initial steps in this direction were driven by the classification of fully gapped and non-interacting fermionic systems according to the Atland-Zirnbauer symmetry classes (combining time-reversal, particle-hole, and chiral symmetries) [37, 38] and in terms of bulk topological invariants such as the $\mathbb{Z}$-valued Chern and winding numbers, or the $\mathbb{Z}_2$-valued Fu-Kane and Chern-Simons invariants [39-42]. In photonics, Chern insulators are $\mathbb{Z}$-valued topological phases of matter rooted in time-reversal symmetry breaking [16-19]. Conversely, topological insulators are typically time-reversal invariant systems where topological effects emerge either from pseudospin conservation ($\mathbb{Z}$-valued spin Chern numbers) [20, 22, 24], or from symmetries which enable bosonic Kramers doublets ($\mathbb{Z}_2$-valued invariants) [8, 43], or both [21, 23, 44, 45].

More recently, emphasis has also been placed on the topological role of the photonic particle-hole symmetry stemming from the reality of the electromagnetic field. This symmetry has been shown to prevent systems from having a ground state, which leads to unique photonic topological phases [45-48].

In photonic crystals, the classifications above can be altered and enriched due to the additional spatial symmetries (discrete translations and point-group transformations) [14, 15, 34-36]. The band Chern numbers in these systems are predictable from the point-group symmetry representations of the electromagnetic fields at high-symmetry points of the FBZ [49]. The impact of crystalline symmetry on topology was first noticed in electronic systems [50-55], opening the possibility to transfer knowledge from fermionic to bosonic systems. For instance, gapped and gapless TRI two-dimensional electronic crystals with lattice gauge fluxes have also been classified according to the projective representations of the corresponding crystal space groups [56, 57]. The photonic counterpart of such an analysis has yet to be done.

In this work, we develop a novel symmetry-based rule for classifying the topological phases of lossless photonic crystals with broken time-reversal symmetry. Our analysis focuses on an electromagnetic analogue of graphene which is modified periodically with nonreciprocal elements. The justification for choosing this framework is two-fold. First, graphene's unique characteristics, which arise from its Dirac cone spectra, are well-understood within topological frameworks [58, 59], making it an ideal system for exploring topological phases. Second, the breaking of time-reversal symmetry (TRS) is essential for lifting the conical degeneracies and engineer Chern gapped phases. When the electrodynamics is conservative, TRS and Lorentz reciprocity are equivalent [60, 61]. Hence, nonreciprocal dopants are required to induce a crystalline Chern phase in photonic graphene.

Photonic analogues of graphene have been realized in different ways, such as by arranging scatterers (dielectric [62-68] or metallic [69-71]) and semiconductor cavities into honeycomb lattices [72]. Here, we build on the photonic graphene introduced in Ref. [73] which is formed by an array of isotropic dielectric rods embedded in a metallic background. The crystal displays a natural set of electromagnetic modes localized around each dielectric rod, which we use to construct a tight-binding model. Specifically, we build a next-nearest neighbor tight-binding model directly from the photonic orbitals with lowest positive frequency (spectrum is discretized), bypassing Wannierization algorithms [74-76]. The resulting tight-binding Hamiltonian has the usual $2\times2$ matrix structure whose diagonal/off-diagonal elements concern intralattice/interlattice hoppings.

We assume that the nonreciprocal response of the periodic inclusions in the modified graphene is weak, so that it can be treated perturbatively. Besides, we demand that it preserves spatial homogeneity along the axes of the dielectric rods, so that transverse electric (TE) and transverse magnetic (TM) modes are decoupled in both the pristine and "nonreciprocally doped" graphenes. This allows us to develop a tight-binding model for the nonreciprocal graphene using the orbitals of the bare crystal. In our framework, the tight-binding Hamiltonian acquires a term which breaks time-reversal symmetry after the nonreciprocal elements are added.

We demonstrate that the most general nonreciprocal couplings compatible with our premises fall into the gyromagnetic, pseudo-Tellegen, and moving medium classes [77]. Furthermore, TE waves do not distinguish between the pseudo-Tellegen and moving medium couplings, and both types of material responses lead to qualitatively similar physical effects. The gyromagnetic and magnetoelectric perturbations to the response dyadics can be expressed in terms of a scalar field $\kappa(\mathbf{r})$ and a vector field $\mathbf{v}(\mathbf{r})$ as $\overline{\delta\mu}=i\kappa\hat{\mathbf{z}}\times\mathbf{1}_{3\times3}$ and $\overline{\delta\xi}=(\hat{\mathbf{z}}\times\mathbf{v})\times\mathbf{1}_{3\times3}$. Our analysis focuses on the impact of the spatial symmetries of $\kappa(\mathbf{r})$ and $\mathbf{v}(\mathbf{r})$ on the system topology. One interesting question arises: What are the necessary conditions for the nonreciprocal fields to ensure that the perturbation induces a nontrivial topology? Is it sufficient to have nontrivial nonreciprocal couplings, or are there additional requirements related to symmetry?

The scalar and vector fields $\kappa(\mathbf{r})$ and $\mathbf{v}(\mathbf{r})$ are defined over the crystal plane and determine two-dimensional patterns whose spatial symmetries must form any of the wallpaper groups. From the total of 17 of these groups, only 8 account for symmetry under translations of the hexagonal lattice. For simplicity, we consider that the gyromagnetic and magnetoelectric fields, $\kappa(\mathbf{r})$ and $\mathbf{v}(\mathbf{r})$, are symmetric under three fold rotations about any node of the honeycomb grid ($C_3$-symmetry). This leaves us with 5 of the 8 wallpaper groups which might model the nonreciprocal elements in our platform. Their IUC designations are *p6m*, *p6*, *p31m*, *p3m1*, and *p3*.

Based on such premises, we analyze how the nonreciprocal fields must break or preserve certain symmetries to open a topologically nontrivial bandgap. Our findings reveal that simply having nonreciprocal couplings is not enough; specific symmetry constraints must be satisfied for the system to achieve a non-zero Chern number, thereby ensuring a topologically protected phase. Our key result is the discovery that the *p6m*-component of the nonreciprocal fields must be nontrivial in order that a crystal with a photonic bandgap can have a non-zero Chern number. We derive this rule by analyzing how the spatial symmetries of the fields $\kappa(\mathbf{r})$ and $\mathbf{v}(\mathbf{r})$ constrain the nonreciprocal lattice hoppings in the tight-binding picture.

The paper is organized as follows. In Sec. II, we characterize the electromagnetic fields of the fundamental photonic mode (orbital) localized around a single dielectric rod. In Sec. III, we lay out the foundations of the tight-binding model built upon the single-rod orbitals. In Sec. IV, we show that the most general nonreciprocal responses that ensure the decoupling between TE and TM waves fall into the classes of gyromagnetic, pseudo-Tellegen, or moving media. In Sec. V, we derive the TRS-breaking correction to the tight-binding Hamiltonian of bare graphene, arising from the perturbative $C_3$-symmetric nonreciprocal response. In Sec. VI, we study how the spatial symmetries of the nonreciprocal response constrain the tight-binding interaction Hamiltonian and, thereby, the topology of the photonic crystal. Most notably, we show that the fields $\kappa(\mathbf{r})$ and $\mathbf{v}(\mathbf{r})$ must have a nonvanishing *p6m*-component so that the spectrum is gapped and topological. In Sec. VII, we compare the results of our tight-binding model and topological analysis against full wave numerical simulations relying on the plane wave method [78]. Finally, in Sect. VIII we present a short summary of the main findings.

## II. Localized Modes of a Single Scatterer

Consider a single circular air rod with radius $R$, embedded in a metallic background [see Fig. 1(a)]. For convenience, we adopt a system of cylindrical coordinates $(\hat{\boldsymbol{\rho}}, \hat{\boldsymbol{\phi}}, \hat{\mathbf{z}})$. Supposing that the metal follows a lossless Drude dispersion model, the relative permittivity of the system is described by:

$$\varepsilon(\rho) = 1 - \frac{\omega_p^2}{\omega^2} \Theta(\rho - R). \tag{1}$$

Here $\Theta$ is the Heaviside step function, $\omega_p$ is the plasma frequency of the metal and $\omega$ is the oscillation frequency (the time variation $e^{-i\omega t}$ is implicit). Due to the cylindrical geometry of the system, the electromagnetic modes of the air scatterer can be split into TE and TM waves. Here, we shall only be concerned with TE photonic excitations, for which the complex electromagnetic fields are of the form $\mathbf{E} = E_z \hat{\mathbf{z}}$ and $\mathbf{H} = H_\rho \hat{\boldsymbol{\rho}} + H_\phi \hat{\boldsymbol{\phi}}$ with $\partial / \partial z = 0$.

In the absence of charge and current sources, the electric field of the TE states must be such that

$$\hat{\mathcal{H}} E_z \equiv \left[ -c^2 \nabla^2 + \omega_p^2(\rho) \right] E_z = \omega^2 E_z, \tag{2}$$

with $c$ the speed of light in vacuum and $\omega_p(\rho) \equiv \omega_p \Theta(\rho - R)$ ($\omega_p(\rho) = 0$ inside the rod and $\omega_p(\rho) = \omega_p$ outside the rod). The nonhomogeneous Helmholtz equation (2) is formally equivalent to the time-independent Schrodinger equation $\left[ -\hbar^2 \nabla^2 / 2m + \mathcal{V}(\rho) \right] \psi = \mathcal{E} \psi$ for a particle of mass $m$, energy $\mathcal{E}$, in a central potential $\mathcal{V}(\rho)$, and wavefunction $\psi$. In particular, the equations match when $\mathcal{V}(\rho) = \hbar^2 \omega_p^2 \Theta(\rho - R)/2mc^2$ and $\mathcal{E} = \hbar^2 \omega^2 / 2mc^2$. This means that the rod-metal system is the photonic analogue of the cylindrical potential well problem in quantum mechanics. Accordingly, the operator $\hat{\mathcal{H}}$ in Eq. (2) has a frequency spectrum with a discrete real-valued part $\omega_n^2$ that is bounded from below. The corresponding set of localized eigenmodes is denoted by $E_{z,n}$. Henceforward, we assume $\omega_n > 0$ and refer to $\omega_n$ as the eigenvalues of Eq. (2) for simplicity. We are interested in the non-degenerate ground state $E_{z,0}$ associated with the smallest frequency $\omega_0$.

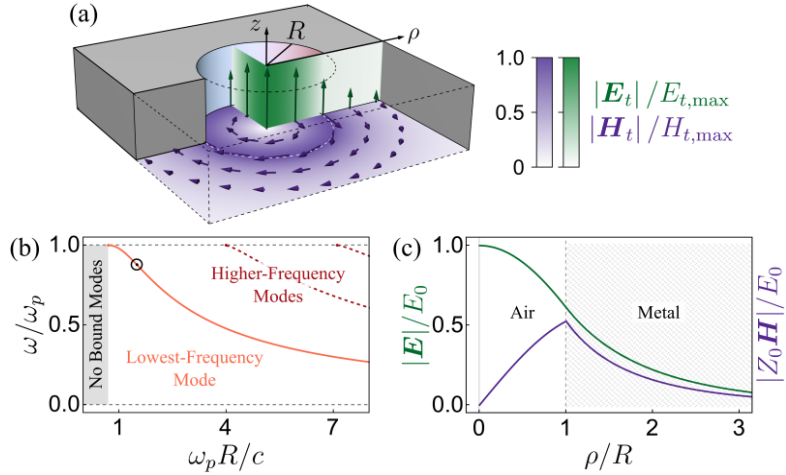

**Fig. 1** (a) model of the circular air rod of radius $R$, embedded in a metallic background. The electric (magnetic) field lines $\mathbf{E}_t$ ($\mathbf{H}_t$) of the TE ground state are represented with green (purple) scaled-down arrows (time snapshots of the fields). In the background, the density plots indicate the magnitude of the fields normalized to the maximum ($\mathbf{E}_{t,\max}$ or $\mathbf{H}_{t,\max}$). (b) Normalized eigenfrequencies $\omega_n$ for bound modes with vanishing angular momentum along the z-axis, as a function of the normalized radius $R$ of the air cylinder. The lowest-order branch gives the ground state frequency $\omega_0$. (c) Magnitudes of the electric and magnetic fields of the TE ground state as a function of the distance $\rho$ to the center of the rod. In (a) and (c), the electromagnetic fields are calculated using $\omega_0 \approx 0.88\omega_p$, for a system with $\omega_p/Rc = 1.5$ [point circled in black in (b)].

Due to rotational invariance, the $z$-component of the orbital angular momentum of light is conserved. At the ground state frequency, the angular momentum should vanish. Therefore, we look for solutions of Eq. (2) that are independent of the azimuthal coordinate $\phi$. Considering that the electric field must be continuous at the boundary between air and metal, we find that

$$E_{z,n}(\rho) = E_0 \begin{cases} J_0(k_n\rho), & \rho \leq R \\ J_0(k_n R) \dfrac{K_0(\kappa_n \rho)}{K_0(\kappa_n R)}, & \rho > R \end{cases} \quad (3)$$

where $k_n \equiv \omega_n/c$ and $\kappa_n \equiv (\omega_p^2 - \omega_n^2)^{1/2}/c$ are real positive propagation constants, $E_0$ is the peak amplitude at the symmetry-axis of the air rod, and $J_m$ ($K_m$) is the $m$-th order standard (modified) Bessel function of the first (second) kind. The corresponding magnetic field is azimuthal, $\mathbf{H}_n = H_{\phi,n}\hat{\boldsymbol{\phi}}$. From Faraday's law, $\partial_\rho E_{z,n} = -i\omega_n \mu_0 H_{\phi,n}$, we have

$$H_{\phi,n}(\rho) = -i\dfrac{E_0}{Z_0} \begin{cases} J_1(k_n \rho), & \rho \leq R \\ \dfrac{\kappa_n}{k_n} J_0(k_n R) \dfrac{K_1(\kappa_n \rho)}{K_0(\kappa_n R)}, & \rho > R \end{cases} \quad (4)$$

where $Z_0$ is the impedance of free space. The continuity of the magnetic field at the air-metal interface yields the secular equation

$$\dfrac{J_1\left(\dfrac{\omega_n R}{c}\right)}{J_0\left(\dfrac{\omega_n R}{c}\right)} = \dfrac{(\omega_p^2 - \omega_n^2)^{\frac{1}{2}}}{\omega_n} \dfrac{K_1\left((\omega_p^2 - \omega_n^2)^{\frac{1}{2}} \dfrac{R}{c}\right)}{K_0\left((\omega_p^2 - \omega_n^2)^{\frac{1}{2}} \dfrac{R}{c}\right)}, \quad (5)$$

which admits countable number of solutions of the form $\omega_n = \omega_n(\omega_p, R)$. These are the eigenfrequencies $\omega_n$ of the bound modes with vanishing angular momentum along the $z$-axis, determined by the physical and geometrical parameters of the dielectric "well". The smallest of all solutions is the ground state frequency $\omega_0$. We refer to this mode as the single-rod ground state orbital throughout the text. In Fig. 1(b), we show that the normalized frequencies $\omega_n/\omega_p$ decrease monotonically as the normalized radius $\omega_p R/c$ increases. This is expected from the analogy with quantum mechanics: the eigenenergies decrease for wells with greater depths or wider apertures. In our system, $\omega_p$ determines the depth and $R$ the well aperture.

The fields $\mathbf{E} = E_{z,0}\hat{\mathbf{z}}$ and $\mathbf{H} = H_{\phi,0}\hat{\boldsymbol{\phi}}$ [given by Eqs. (3) and (4) with $n = 0$] determine the TE "ground state" of the rod-metal system. In Fig. 1(c), we plot the magnitude of the electric and magnetic fields as a function of the distance $\rho$ to the center of the rod. We set $\omega_p R/c = 1.5$, which corresponds to the ground state frequency $\omega_0 \approx 0.88\omega_p$ [circled in black in Fig. 1(b)]. The fields have an evanescent character inside the opaque metal ($\omega_0 < \omega_p$). The electric (magnetic) field peak is reached at the center (perimeter) of the air cylinder. Figure 1(a) displays the field amplitude density plots and the fields vector lines over different cross-sections of the rod. Specifically, we display the instantaneous electric (magnetic) vector field $\mathbf{E}_t$ ($\mathbf{H}_t$) calculated for $t = \pi/4\omega$, with the green (purple) arrows scaled down to fit the model. In the background, the density plots indicate the magnitude of the field, normalized to the maximum ($\mathbf{E}_{t,\max}$ or $\mathbf{H}_{t,\max}$).

## III. Tight-Binding Model of Photonic Graphene

Consider now a honeycomb array of parallel and identical air rods in the metallic background, as illustrated in Fig. 2(a). We study this photonic crystal in Cartesian coordinates $(\hat{\mathbf{x}}, \hat{\mathbf{y}}, \hat{\mathbf{z}})$, where the axes of revolution of the rods are aligned with the $z$-axis and intersect the $xoy$-plane at the nodes of the sublattices $A$ and $B$ with positions $\mathbf{R}_{A,\mathbf{n}} = n_1 \mathbf{a}_1 + n_2 \mathbf{a}_2$ and $\mathbf{R}_{B,\mathbf{n}} = n_1 \mathbf{a}_1 + n_2 \mathbf{a}_2 - \boldsymbol{\delta}_2$. Here, $\mathbf{n} = (n_1, n_2) \in \mathbb{Z}^2$ is a pair of integers that multiply the primitive lattice vectors $\mathbf{a}_1 = 3a\hat{\mathbf{x}}/2 - \sqrt{3}a\hat{\mathbf{y}}/2$ and $\mathbf{a}_2 = 3a\hat{\mathbf{x}}/2 + \sqrt{3}a\hat{\mathbf{y}}/2$, $\boldsymbol{\delta}_2 = -a\hat{\mathbf{x}}$ interchanges sublattices, and $a$ is the spacing between nearest neighbors [see Fig. 2(a)]. We set the radius of the rods equal to $R = 0.3a$, so they do not overlap each other.

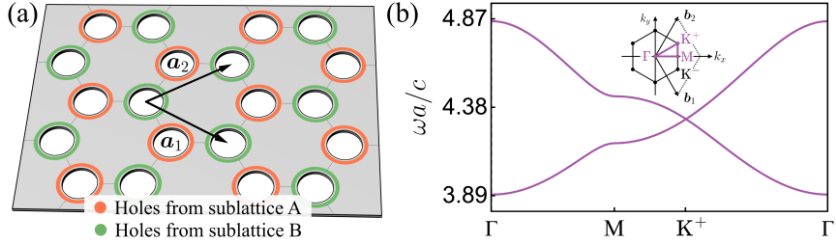

**Fig. 2** Model for the photonic analogue of graphene. A honeycomb array of holes is drilled on a metallic board. The radius of the holes is $R = 0.3a$, with $a$ the nearest neighbors distance. The plasma frequency of the metallic host is set equal to $\omega_p a/c = 5.0$. (b) Band structure of the photonic graphene along the path $\Gamma - M - K^+ - \Gamma$ of the FBZ (shown in the inset).

It was seen in Sec. II that an individual rod in the metallic host acts as a potential well that supports bound eigenmodes with a discrete spectrum. In the multi-well crystal introduced above, the low-frequency photonic states can be approximated by linear combinations of the single-well eigenmodes, if the latter are strongly localized around each rod. This is the standard tight-binding approximation. The length of localization should be smaller than the lattice constant $a$, so that the inter-well hybridization is weak. We can translate this condition to quantitative terms in the case of the TE ground state of the rod-metal system. Indeed, Eqs. (3) and (4) show that the electric and magnetic fields decay radially inside the metal as $E_{z,0} \propto K_0(\kappa_0 \rho)$ and $H_{\phi,0} \propto K_1(\kappa_0 \rho)$. Asymptotically, the modified Bessel functions of the second kind behave as damped exponentials: $K_0(\kappa_0 \rho) \approx K_1(\kappa_0 \rho) \propto e^{-\kappa_0 \rho}$ for $\kappa_0 \rho \gg 1$. Thus, under the condition $\kappa_0 a \gg 1$, the inter-well hybridization is always weak. As already discussed, Figs. 1(a) and 1(c) represent the TE ground state of the single-rod system with $\omega_p R/c = 1.5$, to which corresponds the ground state frequency $\omega_0 \approx 0.88 \omega_p$. From these values and $R = 0.3a$, it is found that $\kappa_0 a = 2.39$. Therefore, for this set of parameters, it is possible to build a reliable tight-binding model of the crystal in Fig. 2(a).

In Appendix A, we present a detailed derivation of the tight-binding model, considering interactions up to second nearest neighbors. The resulting $2\times 2$ tight-binding Hamiltonian $\mathcal{H}_{\mathbf{k}}$ is of the form:

$$\mathcal{H}_{\mathbf{k}} = d_0(\mathbf{k})\, \mathbf{1}_{2\times 2} + \mathbf{d}(\mathbf{k})\cdot\boldsymbol{\sigma}, \tag{6}$$

with $\mathbf{1}_{2\times 2}$ the identity matrix, $\mathbf{d}=(d_x, d_y, d_z)$, $\boldsymbol{\sigma}=(\sigma_x,\sigma_y,\sigma_z)$ is built from the three Pauli matrices, and

$$\begin{cases} d_0(\mathbf{k}) = 1-\Omega-\tilde{t}g(\mathbf{k}) \\ d_x(\mathbf{k}) = -t\,\mathrm{Re}\{f(\mathbf{k})\} \\ d_y(\mathbf{k}) = -t\,\mathrm{Im}\{f(\mathbf{k})\} \\ d_z(\mathbf{k}) = 0 \end{cases} \tag{7}$$

The functions of quasi-momentum are

$$f(\mathbf{k}) = e^{-i\mathbf{k}\cdot\boldsymbol{\delta}_2}\sum_{i=1}^{3} e^{i\mathbf{k}\cdot\boldsymbol{\delta}_i} \quad \text{and} \quad g(\mathbf{k}) = \sum_{i=1}^{6} e^{i\mathbf{k}\cdot\boldsymbol{\gamma}_i}, \tag{8}$$

where $\boldsymbol{\delta}_i$ ($\boldsymbol{\gamma}_i$) is the in-plane vector linking a lattice node to its $i$-th first (second) neighbor [see Fig. 3(a)]. The parameter $\Omega$ is the so-called self-frequency term, while $t$ ($\tilde{t}$) is the hopping between first (second) neighbors [see Fig. 3(b)]. These tight-binding parameters are determined by overlapping integrals between pairs of atomic orbitals weighted by the permittivity of the crystal, i.e., the Drude permittivity. We provide detailed expressions for the overlap integrals in Appendix A. The self-frequency and the hopping parameters are dimensionless, real-valued, positive, and small, i.e., $\Omega, t, \tilde{t} \ll 1$. By construction, the crystal is p6m-symmetric. Specifically, the hexagonal primitive cell is invariant under 6-fold rotations about its center and under reflections about any of its symmetry axes. The function $f$ ($g$) in Eq. (8) is the sum of phase shifts acquired by the wave as it propagates from a crystal node to each of its first (second) neighbors.

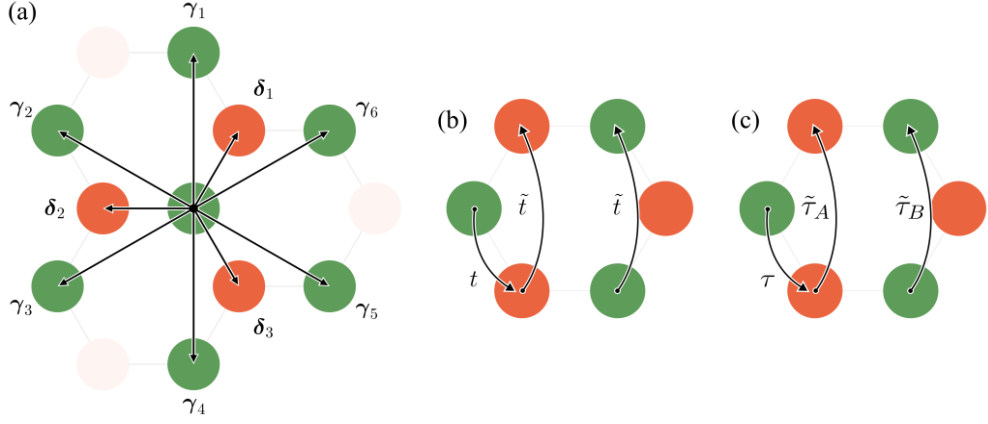

**Fig. 3** (a) The vectors $\boldsymbol{\delta}_i$ ($\boldsymbol{\gamma}_i$) link a generic lattice node to its *i*-th first (second) neighbor. (b) and (c) Illustration of the reciprocal (nonreciprocal) hoppings $t$ and $\tilde{t}$ ($\tau$, $\tilde{\tau}_A$, and $\tilde{\tau}_B$) on the hexagonal unit cell, respectively.

$\mathcal{H}_{\mathbf{k}}$ is the well-known second-neighbors tight-binding Hamiltonian of graphene. It is defined in such a way that its eigenvalues determine the eigenfrequencies of the photonic crystal normalized to the ground state eigenfrequency $\omega_0$ of the single-rod system. From Eqs. (6) and (7), the eigenvalues are simply

$$\omega_{\pm}(\mathbf{k})\Big|_{\substack{\text{bare}\\\text{graphene}}} = \omega_0 \left[ d_0(\mathbf{k}) \pm |t||f(\mathbf{k})| \right]. \tag{9}$$

The spectrum has $C_6$-symmetry about the center of the hexagonal FBZ. Besides, the frequency bands $\omega_-$ and $\omega_+$ coalesce at the zeros of the function $f(\mathbf{k})$, the so-called Dirac points $\mathrm{K}^+$ and $\mathrm{K}^-$ with

$$\mathrm{K}^{\pm} = \frac{2\pi}{3\sqrt{3}a}\left(\sqrt{3},\pm 1\right) \tag{10}$$

in the $k_x o k_y$ plane [see the inset of Fig. 2(b)]. Figure 2(b) displays the two lowest-order positive frequency branches along the closed path $\Gamma - \mathrm{M} - \mathrm{K}^+ - \Gamma$ which crosses one of the Dirac points. The bands are obtained using the plane wave expansion method and confirm the qualitative predictions from the tight-binding model.

# IV. Nonreciprocal perturbations

Next, we examine the nonreciprocal modifications of graphene that are compatible with the propagation of TE waves. A reciprocal and conservative photonic platform has a trivial Chern topology. In fact, Lorentz reciprocity guarantees that light transport is necessarily bidirectional and thus sensitive to back-scattering. Here, we note in passing that scattering anomalies can be engineered in some reciprocal systems if certain duality symmetries are enforced, enabling propagation free of backscattering in a bidirectional system [21, 44]. Notably, these anomalies can be a hallmark of a different class of photonic topological states. Nevertheless, to create a Chern phase, understood as the strongest form of topological protection, one must break time-reversal symmetry. Thus, to engineer a crystalline Chern phase, it is necessary to modify the graphene's response with a nonreciprocal coupling.

For simplicity, we restrict our analysis to nonreciprocal conservative couplings that share the same periodicity as the honeycomb lattice. Additionally, we require that the modified crystal maintains homogeneity along the spatial axis ($z$-direction), and that the decoupling between TE and TM waves is preserved. Maxwell's equations in a generic bianisotropic photonic system take the form

$$ic\begin{pmatrix} \mathbf{0}_{3x3} & \nabla \times \mathbf{1}_{3x3} \\ -\nabla \mathbf{1}_{3x3} & \mathbf{0}_{3x3} \end{pmatrix} \cdot \mathbf{f} = \omega \begin{pmatrix} \overline{\varepsilon} & \overline{\xi} \\ \overline{\zeta} & \overline{\mu} \end{pmatrix} \cdot \mathbf{f}, \qquad (11)$$

where $\mathbf{f} = (\mathbf{E}, \mathbf{H}Z_0)^{\mathrm{T}}$ is a 6-vector containing all components of the electric and magnetic fields. Here, $\overline{\varepsilon}$, $\overline{\mu}$, $\overline{\xi}$, and $\overline{\zeta}$ are $3\times 3$ tensors which determine the electromagnetic response of the materials. The first two tensors are the relative permittivity and permeability, respectively. The other two determine the magnetoelectric coupling. A material is lossless when

$$\overline{\varepsilon} = \overline{\varepsilon}^{\dagger}, \ \overline{\mu} = \overline{\mu}^{\dagger}, \ \overline{\xi} = \overline{\zeta}^{\dagger}. \qquad (12)$$

On the other hand, a reciprocal response requires that

$$\overline{\varepsilon} = \overline{\varepsilon}^{\mathrm{T}}, \quad \overline{\mu} = \overline{\mu}^{\mathrm{T}}, \quad \overline{\xi} = -\overline{\zeta}^{\mathrm{T}}. \tag{13}$$

We focus on the nonreciprocal components of the perturbation defined such that $\overline{\delta\varepsilon} = -\overline{\delta\varepsilon}^{\mathrm{T}}$, $\overline{\delta\mu} = -\overline{\delta\mu}^{\mathrm{T}}$, $\overline{\delta\xi} = \overline{\delta\zeta}^{\mathrm{T}}$. If the nonreciprocal perturbations are subject to the constraint in Eq. (12), so that they are consistent with a conservative response, it follows that $\overline{\delta\varepsilon}$ and $\overline{\delta\mu}$ ($\overline{\delta\xi}$ and $\overline{\delta\zeta}$) must be purely imaginary (real) tensors. If the material is also spatially homogeneous along the $z$-direction and does not mix between TE and TM modes, some elements of the tensors must vanish. Specifically, the nonreciprocal (and conservative) perturbations of the permittivity and permeability must be of the form

$$\overline{\delta\varepsilon} = i\begin{pmatrix} 0 & \kappa_\epsilon & 0 \\ -\kappa_\epsilon & 0 & 0 \\ 0 & 0 & 0 \end{pmatrix}, \quad \overline{\delta\mu} = i\begin{pmatrix} 0 & \kappa_\mu & 0 \\ -\kappa_\mu & 0 & 0 \\ 0 & 0 & 0 \end{pmatrix} \tag{14}$$

with $\kappa_\epsilon, \kappa_\mu$ real-valued. This means that the electric-electric and magnetic-magnetic nonreciprocal couplings can only be gyrotropic. Conversely, the magnetoelectric nonreciprocal (and conservative) perturbations need to be structured as

$$\overline{\delta\xi} = \begin{pmatrix} 0 & 0 & w_x \\ 0 & 0 & w_y \\ v_x & v_y & 0 \end{pmatrix}, \tag{15}$$

with $\overline{\delta\zeta} = \overline{\delta\xi}^{\mathrm{T}}$, with $v_x, v_y, w_x, w_y$, real-valued numbers. In words, the nonreciprocal perturbation of the magnetoelectric response can be any mixture of the pseudo-Tellegen ($\overline{\delta\xi}$ symmetric) and moving medium ($\overline{\delta\xi}$ anti-symmetric) classes [77].

For the nonreciprocal perturbations defined in Eqs. (14) and (15), the Maxwell's equations for TE waves in the modified photonic graphene reduce to

$$\mathcal{D}\mathbf{\psi} = \omega\left[\mathcal{M}_c + \begin{pmatrix} 0 & v_x & v_y \\ v_x & 0 & i\kappa_\mu \\ v_y & -i\kappa_\mu & 0 \end{pmatrix}\right] \cdot \mathbf{\psi} \tag{16}$$

Here, $\mathcal{D}$ is the differential operator defined in Eq. (A2b) of Appendix A, $\mathcal{M}_c$ is the material matrix of the pristine (unperturbed) graphene, and $\boldsymbol{\psi} = (E_z, H_x Z_0, H_z Z_0)^\mathrm{T}$ is the state vector. The above result reveals two key points. First, the TE modes are insensitive to the gyroelectric response. This is because the electric field is aligned with the axis of gyration ($z$-axis) and is thus unaffected by this type of perturbation. Second, TE modes are only influenced by half of the elements in the magnetoelectric tensors, namely, the elements $v_x$ and $v_y$. This means that TE waves cannot distinguish between the pseudo-Tellegen and moving medium responses, which can then be described in terms of a single pseudo-vector $\mathbf{v} \equiv v_x \hat{\mathbf{x}} + v_y \hat{\mathbf{y}}$. In accordance, the nonreciprocal perturbations of the bare photonic graphene can always be described by scalar and vector fields, $\kappa = \kappa(\mathbf{r})$ and $\mathbf{v} = \mathbf{v}(\mathbf{r})$, respectively, with the periodicity of the honeycomb grid. We refer to them as the gyromagnetic and magnetoelectric fields that determine the nonreciprocal perturbation of graphene. Figure 4(a) illustrates a generic spatial distribution of the two nonreciprocal fields.

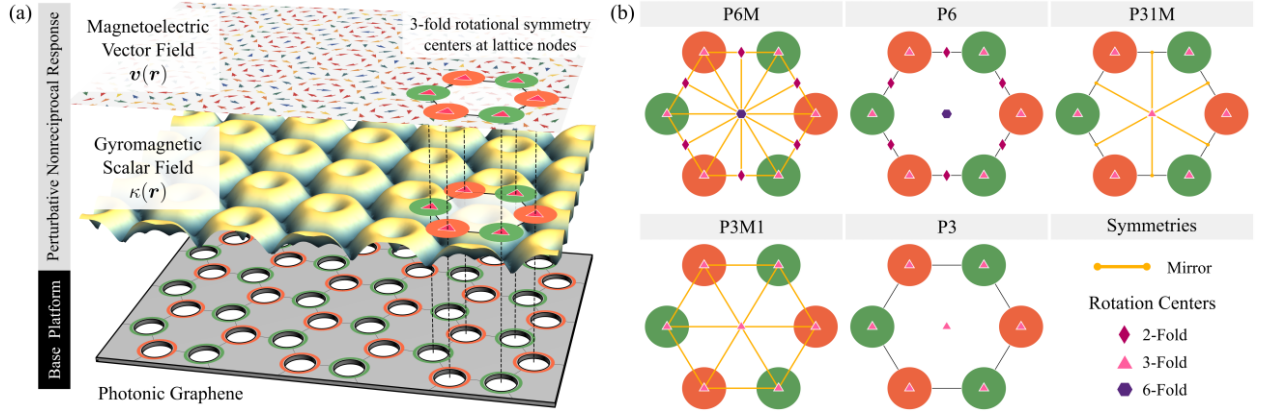

**Fig. 4** (a) Nonreciprocal modification of the photonic graphene. Under the constraints of energy conservation and polarization decoupling, the electromagnetic response acquires gyromagnetic and magnetoelectric components determined by the scalar and vector fields $\kappa(\mathbf{r})$ and $\mathbf{v}(\mathbf{r})$, respectively. (b) All possible wallpaper symmetry groups of the nonreciprocal fields that remain invariant under 3-fold rotations about any node of the honeycomb lattice. Their IUC designations are shown at the top of the unit cells where the mirrors and rotational symmetry centers are marked according to the legend at the bottom right corner.

# V. Tight-Binding Model for the Nonreciprocal Graphene

In Appendix B, we derive a tight-binding model for the photonic graphene with a generic weak nonreciprocal perturbation of the gyromagnetic and/or magnetoelectric type. The tight-binding Hamiltonian of the perturbed crystal is of the form,

$$\mathcal{H}'_{\mathbf{k}} = \mathcal{H}_{\mathbf{k}} - \mathcal{P}_{\mathbf{k}} \tag{17}$$

It is the sum of the Hamiltonian of bare graphene ($\mathcal{H}_{\mathbf{k}}$) corrected by a perturbation term $\mathcal{P}_{\mathbf{k}}$ that accounts for nonreciprocal interactions between "atomic" orbitals. For simplicity, we assume that all nodes of the honeycomb array are 3-fold rotational symmetry centers of the gyromagnetic and magnetoelectric fields, $\kappa(\mathbf{r})$ and $\mathbf{v}(\mathbf{r})$ ($C_3$-symmetry). This spatial symmetry is inherent to the bare crystal structure and plays a critical role in the formation of the Dirac cones. Given this assumption, the symmetry properties of the nonreciprocal fields can be classified by one of five wallpaper groups, represented in Fig. 4(b). Importantly, by preserving the 3-fold rotational symmetry, the matrix structure of the correction term becomes rather simple. Specifically, $\mathcal{P}_{\mathbf{k}}$ can be written in terms of Pauli matrices and of the identity matrix as follows,

$$\mathcal{P}_{\mathbf{k}} = p_0(\mathbf{k})\mathbf{1}_{2\times 2} + \mathbf{p}(\mathbf{k})\cdot\boldsymbol{\sigma}, \tag{18}$$

being the relevant coefficients

$$\begin{cases} p_0(\mathbf{k}) = sh(\mathbf{k}) \\ p_x(\mathbf{k}) = -\tau\,\mathrm{Im}\{f(\mathbf{k})\} \\ p_y(\mathbf{k}) = +\tau\,\mathrm{Re}\{f(\mathbf{k})\} \\ p_z(\mathbf{k}) = mh(\mathbf{k}) \end{cases} \tag{19}$$

with

$$h(\mathbf{k}) = 2\sum_{i=2,4,6} \sin(\mathbf{k}\cdot\boldsymbol{\gamma}_i). \tag{20}$$

The parameter $\tau$ is the nonreciprocal hopping between first neighbors. Conversely, $s$ and $m$ are tight-binding parameters determined by the nonreciprocal hoppings between second neighbors in sublattices $A$ and $B$, $\tilde{\tau}_A$ and $\tilde{\tau}_B$ [see Fig. 3(c)]. Namely,

$$s = \frac{\tilde{\tau}_A + \tilde{\tau}_B}{2} \quad \text{and} \quad m = \frac{\tilde{\tau}_A - \tilde{\tau}_B}{2}. \tag{21}$$

We use the letter "$\tau$" to label the nonreciprocal hoppings that are counterparts of the reciprocal hoppings denoted with the letter "$t$". In both cases, the tilde symbol serves to distinguish between first and second neighbor hoppings. All the hopping parameters, $\tau$, $\tilde{\tau}_A$, and $\tilde{\tau}_B$ are dimensionless and real-valued. Their sign depends on the gyromagnetic and magnetoelectric fields, as can be seen from Eqs. (B11) and (B12) in Appendix B.

In the construction of the tight-binding Hamiltonian, the frequency $\omega$ is set identical to the "atomic" eigenfrequency $\omega_0$; in this manner, the photonic crystal spectrum can be obtained by standard diagonalization of the relevant Hamiltonian. Furthermore, the perturbation of the tight-binding Hamiltonian assumes a weak nonreciprocal interaction. For strong nonreciprocal fields, the strength of the hoppings may be inaccurate and other interactions may be relevant. These approximations are discussed in greater detail in Appendix A and Appendix B.

The photonic crystal generally loses the p6m-symmetry after the introduction of the nonreciprocal perturbation. However, the preservation of $C_3$-symmetry ensures that the nonreciprocal hopping between first neighbors is independent of their relative position. Therefore, for the inter-lattice interactions, $\mathcal{P}_\mathbf{k}$ depends solely on a single degenerate hopping $\tau$. This mirrors the behavior of the bare graphene Hamiltonian $\mathcal{H}_\mathbf{k}$, which depends solely on $t$, its reciprocal counterpart. Due to the lower symmetry, the hopping between second neighbors depends on the sublattice, so, in general, $\tilde{\tau}_A \neq \tilde{\tau}_B$. Moreover, these hoppings also depend on the relative position between the second neighbors: a sign change occurs each time we move from neighbor $i$ to the neighbor $i+1$. This sign change is accounted for by the function $h(\mathbf{k})$ in Eq. (20), so that $\mathcal{P}_\mathbf{k}$ only depends on the two coefficients $\tilde{\tau}_A$ and $\tilde{\tau}_B$.

As is well known, the Hamiltonian of bare graphene $\mathcal{H}_\mathbf{k}$ has no $\sigma_z$-component, i.e., $d_z(\mathbf{k}) = 0$ [Eq. (7)]. As a result, the frequency gap between the branches $\omega_+$ and $\omega_-$ is only controlled by $|f(\mathbf{k})|$ which vanishes at the Dirac points [see Eqs. (9) and (10)]. From Eq. (19), one sees that the nonreciprocal perturbation originates a term $m h(\mathbf{k}) \sigma_z$ which may lift the Dirac degeneracies. To see how, consider the eigenvalues of $\mathcal{H}'_\mathbf{k}$ which are the frequency bands $\omega'_\pm(\mathbf{k})/\omega_0 = d'_0(\mathbf{k}) \pm |\mathbf{d}'(\mathbf{k})|$, with $d'_0 = d_0 - p_0$ and $\mathbf{d}' = \mathbf{d} - \mathbf{p}$. They can be written more explicitly as

$$\omega'_\pm(\mathbf{k}) = \omega_0 \left[ d'_0(\mathbf{k}) \pm \sqrt{|T|^2 |f(\mathbf{k})|^2 + m^2 h^2(\mathbf{k})} \right], \qquad (22)$$

with $T = t + i\tau$. The above result establishes that the bands $\omega'_+$ and $\omega'_-$ can only coalesce at the simultaneous zeros of $f(\mathbf{k})$ and $h(\mathbf{k})$, i.e., if $h(\mathbf{k})$ vanishes at the Dirac points. But since $h(\mathrm{K}^\pm) = \pm 3\sqrt{3}$, it follows that the spectrum is necessarily gapped when $m$ is nontrivial. We refer to $m$ as the "split" parameter of the Hamiltonian $\mathcal{H}'_\mathbf{k}$ because it controls the split of the Dirac degeneracies and opening of the photonic bandgap. It is determined by the (half) difference of the nonreciprocal hoppings between second neighbors [see Eq. (21)]. On the other hand, $s$ is the average of those hoppings and determines an asymmetric shift of the two Dirac cones in frequency. We shall refer to it as the "shift" parameter.

Because the nonreciprocal perturbation preserves the $C_3$ node-symmetry, $p_{x,y}$ is governed by the same function $f(\mathbf{k})$ that determines $d_{x,y}$ [Eq. (7)]. This property allows us to define the complex-valued hopping $T$ in Eq. (22). In particular, when $m = 0$, the frequency bands of the modified graphene still form Dirac cones centered at $\mathrm{K}^+$ and $\mathrm{K}^-$. The photonic Fermi velocity $v_F$ changes when $\tau \neq 0$. For the original crystal $v_F = 3at/2$, while $v_F = 3a\sqrt{t^2 + \tau^2}/2$ after the perturbation.

## VI. Symmetry Decomposition of the Nonreciprocal Fields and Topology

The hopping parameters $\tilde{\tau}_A$ and $\tilde{\tau}_B$ are determined by the overlap between "atomic" orbitals weighted by the nonreciprocal response of the crystal. In turn, this response is governed by the nonreciprocal fields $\kappa(x, y)$ and $\mathbf{v}(x, y)$. It is useful to decompose $\kappa$ and $\mathbf{v}$ into components with particular spatial-symmetries which contribute differently to the hoppings $\tilde{\tau}_A$ and $\tilde{\tau}_B$, and thus to the split parameter $m$. Here, we identify the symmetry classes that control the opening of the photonic bandgap (corresponding to $m \neq 0$) and other spectral effects.

In our analysis, the gyromagnetic and magnetoelectric fields are treated perturbatively ($\kappa, |\mathbf{v}| \ll 1$), but otherwise may be arbitrarily defined over the unit cell. Let $M_x$ and $M_y$ be the mirror axes of the hexagonal unit cell of the honeycomb grid [see Fig. 5(a) in Sec. VII]. The nonreciprocal couplings can be decomposed into components which have well-defined parities under reflections about those axes. Specifically, we can write

$$\begin{cases} \kappa = \kappa^{--} + \kappa^{-+} + \kappa^{+-} + \kappa^{++} \\ \mathbf{v} = \mathbf{v}^{--} + \mathbf{v}^{-+} + \mathbf{v}^{+-} + \mathbf{v}^{++} \end{cases}, \quad (23)$$

where the elements $\kappa^{p_x p_y}$ and $\mathbf{v}^{p_x p_y}$ are odd (even) under a reflection about the axis $M_{x,y}$ for $p_{x,y} = -1$ ($p_{x,y} = +1$). For a perturbation of the magnetoelectric type, the even and odd parities also account for the vector nature of the field $\mathbf{v}$. The elements are denoted by odd-odd, odd-even, even-odd, or even-even, where the first (second) label determines the parity of the nonreciprocal field under reflection about the $M_x$ ($M_y$) axis. The decompositions in Eq. (23) induce similar splitting in the second neighbor hoppings:

$$\tilde{\tau}_{A,B} = \tilde{\tau}_{A,B}^{--} + \tilde{\tau}_{A,B}^{-+} + \tilde{\tau}_{A,B}^{+-} + \tilde{\tau}_{A,B}^{++}. \quad (24)$$

For example, $\tilde{\tau}_A^{+-}$ is the part of the hopping $\tilde{\tau}_A$ that originates exclusively from the even-odd components $\kappa^{+-}$ and $\mathbf{v}^{+-}$ of the nonreciprocal fields. Some algebra reveals that [see Appendix C]

$$\tilde{\tau}_A^{-\pm} = \tilde{\tau}_B^{-\pm} = 0, \tag{25}$$

i.e., the hopping between second neighbors vanishes if the fields $\kappa$ and $\mathbf{v}$ are odd under an $M_x$-reflection, regardless of the sublattice. Conversely, when the nonreciprocal couplings are even under an $M_x$-reflection, one has

$$\tilde{\tau}_B^{+\pm} = \mp \tilde{\tau}_A^{+\pm}. \tag{26}$$

In this case, the hopping parameters for the two sublattices are additive-symmetric (identical) if the nonreciprocal perturbations are even (odd) under an $M_y$-reflection.

By combining Eqs. (24)-(26) with Eq. (21), we find that the shift ($s$) and split ($m$) parameters can be expressed as:

$$s = \tilde{\tau}_A^{+-} \quad \text{and} \quad m = \tilde{\tau}_A^{++}. \tag{27}$$

Thus, the shift parameter $s$ originates from the even-odd components of the nonreciprocal fields, while the split $m$ is determined by the even-even parts. We recall that the $m$ parameter controls the gap width, whereas $s$ determines an asymmetric frequency shift of the Dirac cones.

Let us now consider nonreciprocal perturbations with well-defined reflection parities, i.e., with a single nontrivial component. The tight-binding perturbation ($\mathcal{P}_\mathbf{k}$) of bare graphene due to the nonreciprocal fields [Eqs. (18) and (19)] is governed by $s$ and $m$, in conjunction with $\tau$. From the symmetry decomposition of $s$ and $m$ [Eq. (27)], one can identify three relevant cases.

The first case arises when the nonreciprocal fields are odd under reflection about $M_x$. In this scenario, the frequency bands form degenerate Dirac cones (around $\text{K}^+$ and $\text{K}^-$) as in bare graphene ($m = 0$), with the Dirac frequency remaining unchanged ($s = 0$). In the second case, when the nonreciprocal fields have even-odd symmetry, the perturbation term has a component

-20-

of the form $\tilde{\tau}_A h(\mathbf{k})\mathbf{1}_{2\times 2}$, with $h(\mathbf{k})$ an odd function over the FBZ and $\tilde{\tau}_B = \tilde{\tau}_A$ ($m = 0, s \neq 0$). This configuration maintains the closure of the Dirac cones but induces an asymmetric shift in their frequencies. In the final case, for the even-even parity class, the perturbation includes a split term $\tilde{\tau}_A h(\mathbf{k})\sigma_z$ that lifts the Dirac degeneracies, opening a spectral gap ($m \neq 0, s = 0$). In this case, $\tilde{\tau}_B = -\tilde{\tau}_A$. Because the perturbation breaks time-reversal symmetry, the photonic band gap may be topological.

The tight-binding perturbation $\mathcal{P}_\mathbf{k}$ also features terms that depend on the inter-lattice interactions, controlled by the nonreciprocal hopping parameter $\tau$ between first neighbors. This hopping is nontrivial only when the fields $\kappa$ and $\mathbf{v}$ have odd-even parity [see Appendix C]. As discussed above, for this symmetry $m = 0 = s$, and thereby the Dirac cones remain closed and centered about the same frequency. The effect of $\tau \neq 0$ is to modify the frequency dispersion slope, corresponding to an increase of the photonic Fermi velocity $v_F$ [see the discussion near Eq. (22)].

As is well-known, the Hamiltonian of bare graphene is time-reversal invariant: $\mathcal{H}_\mathbf{k} = \mathcal{H}_{-\mathbf{k}}^*$. Evidently, the interaction term originating from the nonreciprocal perturbation does not share this symmetry: $\mathcal{P}_\mathbf{k} = -\mathcal{P}_{-\mathbf{k}}^*$. This means that, if the split term is nonvanishing, it takes on opposite signs at each Dirac point. For that reason, if the perturbed crystal has a photonic bandgap, the gap Chern number is necessarily nontrivial. In Appendix D, we prove that

$$\mathcal{C}_{\text{gap}} = \text{sgn}\left(\tilde{\tau}_A^{++}\right), \tag{28}$$

where $\tilde{\tau}_A^{++}$ is the even-even part of the hopping from an "atomic" orbital in sublattice $A$ to its second neighbor directed along $\gamma_1$. Recall that $\tilde{\tau}_A$ measures the overlap between two atomic orbitals over the crystal plane, weighted by the nonreciprocal fields, $\kappa$ and $\mathbf{v}$. If these fields are nonvanishing only in specific regions of the unit cell, the gap Chern number $\mathcal{C}_{\text{gap}}$ can be easily calculated from the integral expressions for the hopping $\tilde{\tau}_A$ [Eqs. (B11) and (B12)] using Eq. (28). The combination of $C_3$-symmetry with the mirrors $M_x$ and $M_y$ determines that the even-even parts of the nonreciprocal fields belong to the wallpaper group *p6m* [see Fig. 4(b)]. Hence, we conclude that a topological gap exists if and only if the nonreciprocal perturbation of graphene has a nonvanishing *p6m*-symmetric component.

So far, our analysis is focused on the nonreciprocal perturbations of photonic graphene and how they affect the topological phases. However, it is also relevant to discuss the effect of reciprocal perturbations. For example, consider that the permittivity of the dielectric rods is slightly altered depending on the sublattice. This reciprocal modification of the crystal can be treated perturbatively as before. As is well known, the fluctuation of the on-site permittivity originates a correction to the tight-binding Hamiltonian of the form $\left[\delta\varepsilon + \delta\tilde{t}_\varepsilon g(\mathbf{k})\right]\sigma_z$, with $\delta\varepsilon$ and $\delta\tilde{t}_\varepsilon$ real-valued parameters. This correction breaks inversion symmetry but not TRS, and so, by itself it cannot create a nontrivial topology. However, it can open a photonic bandgap with a zero Chern number.

Now consider the combined effects of the reciprocal and nonreciprocal perturbations. For weak perturbations, the two perturbations are combined additively. In this case, the nonreciprocal graphene with perturbed on-site permittivities has a tight-binding description analogous to the Haldane model [79, 80]. In fact, the reciprocal term $\left[\delta\varepsilon + \delta\tilde{t}_\varepsilon g(\mathbf{k})\right]\sigma_z$ acts as a mass term, while the nonreciprocal term $mh(\mathbf{k})\sigma_z$ originates an effective magnetic flux. The key distinction between these two terms is related to the parity of the functions $g(\mathbf{k})$ and $h(\mathbf{k})$, which are even and odd, respectively. Both terms contribute to opening a photonic bandgap, but they compete to determine whether the gap is topological or not. For a photonic crystal with a nontrivial mass term, the gap opens even when the *p6m*-symmetric part of the nonreciprocal coupling vanishes, but is guaranteed to be trivial.

# VII. Case Study: Nonreciprocal Inclusion Rings

To illustrate our theory, we consider a specific perturbation of the photonic graphene with nonreciprocal elements. The perturbation consists of a ring of 12 evenly spaced circular inclusions per unit cell, as depicted in Fig. 5(a). Each nonreciprocal element has a radius of $R_i \approx 0.12a$ and is at a distance $d \approx 0.47a$ from the center of the cell. Our geometry is such that the inclusions do not overlap or touch each other and are embedded in the metallic host. Specifically, aside from the nonreciprocal modification, the inclusions have the same optical response as the metal. This ensures that the Drude-dispersion from the photonic graphene is preserved, simplifying the tight-binding analysis. Figures 5(b) and 5(c) display four graphenes modified with gyromagnetic and magnetoelectric elements, respectively. For simplicity, the two types of nonreciprocal responses are considered separately. In all cases, the inclusions are homogeneous, meaning that their electromagnetic response is independent of the observation point within them.

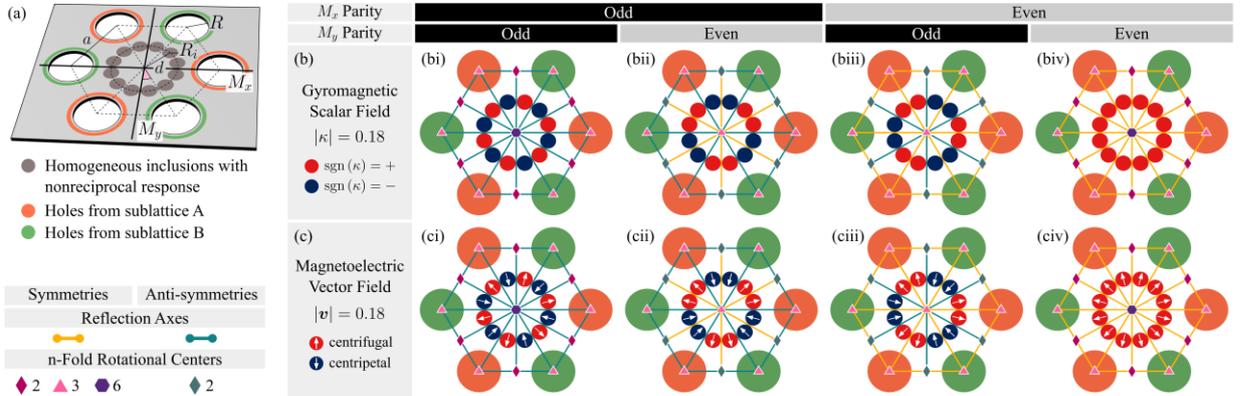

**Fig. 5** (a) Unit cell geometry of the photonic graphene modified with the nonreciprocal elements. The metallic background hosts 12 nonreciprocal circular inclusions forming a ring that is concentric with the cell. The nonreciprocal inclusions are located at a distance of $d = 0.47\,a$ from the center of the cell and have radius $R_i = 0.12\,a$. (b, c) Ensemble of models for which the nonreciprocal response is determined by a field with well-defined parity (even or odd) under reflections about the axes $M_x$ and $M_y$ [see panel (a)]. The nonreciprocal response is of the gyromagnetic/magnetoelectric type for the models in (b)/(c). The gyromagnetic and magnetoelectric nonreciprocal fields, $\kappa$ and $\mathbf{v}$ are defined as in Eqs. (29) and (30) for all 4 parity classes. Their peak magnitudes are $\kappa_m = v_m = 0.18$. The bare graphene is characterized by $\omega_p a/c = 5.0$ and $R/a = 0.3$. All the rotational symmetry and anti-symmetry centers are marked over the unit cells of the nonreciprocal models, according to the legend in the bottom left corner. An anti-symmetry switches the sign of fields $\kappa$ and $\mathbf{v}$ after the mirror transformation. Mirrors and anti-mirrors (reflection axes) are also indicated.

For the gyromagnetic crystals, the nonreciprocal elements have either $\kappa = +0.18$ or $\kappa = -0.18$ [see Fig. 5(b)]. The sign of $\kappa$ varies around the center of the unit cell according to the four parity classes of the gyromagnetic field with respect to reflections about the axes $M_x$ and $M_y$ [see Fig. 5(a)]. Accordingly, we consider four rings of inclusions each corresponding to a gyromagnetic field that belongs to one of the parity classes: odd-odd, odd-even, even-odd, or even-even. As before, the first (second) label identifies the parity of the nonreciprocal field under reflections about the $M_x$ ($M_y$) axis. The gyromagnetic field of the nonreciprocal crystal can be written in a compact form as:

$$\kappa^{p_x p_y}\bigg|_{\text{u.c.}}(\mathbf{r}) = \kappa_m \sum_{n=1}^{12} \Theta(R_i - |\mathbf{R}_n - \mathbf{r}|) s_n^{p_x p_y}. \tag{29}$$

The symbol $|_{\text{u.c.}}$ denotes the restriction of the nonreciprocal field to the unit cell whose center is made coincident with the origin of the $xoy$-plane for convenience. The nonreciprocal inclusion centers are $\mathbf{R}_n = d(\cos\theta_n, \sin\theta_n)$ with the angular positions given by $\theta_n = (2n-1)\pi/12$ ($n=1,\ldots,12$). $s_n^{p_x p_y}$ is the algebraic sign of the gyromagnetic field inside the $n$-th inclusion, that depends on the reflection parities. We use $p_{x,y} = +1/-1$ for even/odd parity under reflection about the axis $M_{x,y}$. For instance, the even-odd response has $s_n^{+-} = (-1)^{\lfloor n/2 \rfloor}$, where $\lfloor x \rfloor$ is the floor function of $x$. The remaining sign functions are $s_n^{++} = 1$, $s_n^{-+} = (-1)^{\lfloor (n-1)/2 \rfloor}$, and $s_n^{--} = (-1)^{n+1}$. The parameter $\kappa_m$ gives the absolute value of the field inside each inclusion and is set equal to $\kappa_m = 0.18$.

The magnetoelectric photonic crystals are constructed in a similar manner. The nonreciprocal response of the inclusions is determined by a pseudo-vector field $\mathbf{v} = v_x\hat{\mathbf{x}} + v_y\hat{\mathbf{y}}$ belonging to one of the well-defined parity classes discussed above [see Fig. 5(c)]. We fix the magnitude of the magnetoelectric vector at $|\mathbf{v}| = 0.18$ and choose it centripetal or centrifugal depending on the specific inclusion. Specifically, $\mathbf{v}$ is constant inside each element and is directed along the radial line that joins the center of the element and the center of the hexagonal cell. The magnetoelectric fields can be written explicitly as:

$$\mathbf{v}^{p_x p_y}\bigg|_{\text{u.c.}}(\mathbf{r}) = v_m \sum_{n=1}^{12} \Theta(|\mathbf{R}_n - \mathbf{r}| - R_i) s_n^{p_x p_y} (\cos\theta_n, \sin\theta_n), \tag{30}$$

where $s_n^{p_x p_y} = +/-$ indicates whether the field is centrifugal or centripetal, and $v_m = 0.18$ is the fixed magnitude of the magnetoelectric vectors (inside the inclusions).

Using the plane wave expansion method [78], we numerically computed the band structures of the nonreciprocal crystals described earlier. Figure 6 displays the photonic dispersion in the vicinity of the Dirac points $K^+$ and $K^-$, along the $k_x$-direction. For comparison, graphene's spectrum is also presented. We focus on the regions surrounding the Dirac points, where the most significant differences between the band structures of the modified and bare graphenes occur.

The results of the simulations confirm the symmetry properties discussed previously. Let us first consider crystals with gyromagnetic elements [Fig. 5(a)]. When the field $\kappa(\mathbf{r})$ is odd under a reflection about the horizontal axis $M_x$, the frequency bands consist of degenerate Dirac cones (around $K^+$ and $K^-$), as in bare graphene [see Figs. 6(*ai*) and 6 (*aii*)]. Note that this is true regardless whether the field exhibits even or odd symmetry under reflection about the (vertical) axis $M_y$. As the nonreciprocal interactions between second neighbors vanish when the field $\kappa$ has odd $M_x$-parity, the dispersion is linear, similar to bare graphene. Indeed, in this case there is no split term to open a gap, and the $C_3$-symmetry of the lattice nodes guarantees the existence of Dirac cones.

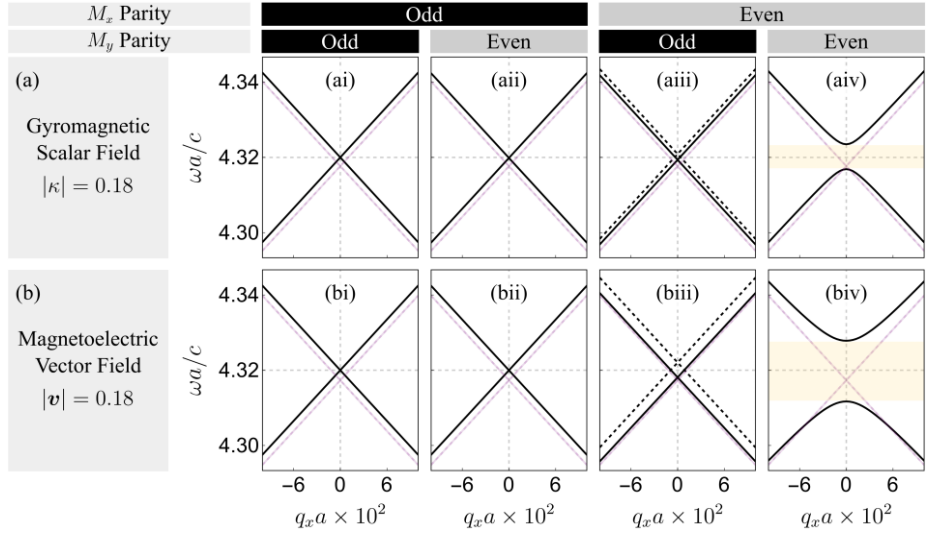

**Fig. 6** - Exact band structures (black lines) of the nonreciprocal crystals whose unit cells are schematized in Fig. 5. For comparison, the bare graphene's bands are also displayed (light pink lines). The frequency dispersion is shown in the vicinity of the Dirac points $K^+$ (solid lines) and $K^-$ (dashed lines), along the $k_x$-direction. The effects of the nonreciprocal perturbations relate consistently to the symmetry class of nonreciprocal field, be it gyromagnetic (top row) or magnetoelectric (bottom row). (a*i*, a*ii*, b*i*, b*ii*) Band structures for odd perturbations under $M_x$-reflection. In this case, the frequency bands form degenerate Dirac cones (solid and dashed lines overlap), as in bare graphene. (a*iii*, b*iii*) Band structures for even-odd perturbations (under reflections about $M_x$-$M_y$, respectively). The linear dispersion is preserved and the two Dirac cones are asymmetrically shifted in frequency. (a*iv*, b*iv*) Band structures for even-even perturbations. The band structure now exhibits a band gap (beige strips).

The linear dispersion is also preserved for nonreciprocal fields belonging to the even-odd class, but in this case the Dirac cones are asymmetrically shifted in frequency between $K^+$ and $K^-$ [see Fig. 6(a*iii*)]. On the contrary, an even-even nonreciprocal perturbation preserves the symmetry of the band structure, ensuring that the dispersions in the vicinity of the Dirac points remain mirror images of each other, but it also results in the opening of a spectral gap [beige strip in Fig. 6(a*iv*)]. These properties align precisely with the predictions of our tight-binding model. An even gyromagnetic field $\kappa$ under reflection about the $M_x$ axis, is essential for enabling nonreciprocal interactions between second neighbors. These interactions give rise to either a shift ($\tilde{\tau}_A h(\mathbf{k})\mathbf{1}_{2\times 2}$) or split ($\tilde{\tau}_A h(\mathbf{k})\sigma_z$) term in the Hamiltonian, depending on the parity of the fields about the $M_y$-axis. Because $h(\mathbf{k})$ is an odd function of the quasimomentum [see Eq. (20)], the shift of the Dirac cones is asymmetric for the even-odd symmetry class.

Curiously, the photonic Fermi velocity $v_F$ is essentially constant across the odd-odd, odd-even, and even-odd parity classes. As previously noted, the nonreciprocal hopping $\tau$ between first neighbors vanishes identically for all parity classes, except the odd-even type. The hopping $\tau$ controls the slope of the Dirac cones: $v_F \propto \sqrt{t^2 + \tau^2}$. The justification for the insensitivity of the Fermi velocity to the nonreciprocal perturbation is that the ratio between the two hopping parameters is quite small: $|\tau/t| \approx 9.0\times 10^{-3}$ ($|\tau/t| \approx 4.0\times 10^{-2}$) for the gyromagnetic (magnetoelectric) perturbation. As a result, to detect the variation in slope caused by $\tau$, one would need to move far away from the Dirac point, reaching a region where the dispersion is no longer linear. The parameter $|\tau/t|$ was numerically estimated by evaluating the hopping integrals [Eqs. (A23), (B11), and (B12)].

Comparing the top and bottom rows of Fig. 6, it is evident that the features of the band structure for different nonreciprocal crystals consistently relate to the parity class of the nonreciprocal field, regardless of whether it is gyromagnetic or magnetoelectric.

For nonreciprocal fields belonging to the even-even symmetry class, the spectrum of the nonreciprocal crystals is gapped (Fig. 6, last column). We have computed the corresponding gap Chern number employing a first principles numerical algorithm relying on photonic Green's functions [81, 82].The results for the two types of nonreciprocal couplings are presented in Figs. 7(ai) and 7(bi) as insets of the band structures near the Dirac points, together with the corresponding unit cell schemes. The gap Chern number of the even-even gyromagnetic (magnetoelectric) model is $\mathcal{C}_{\text{gap}}^{\text{gyr}} = -1$ ($\mathcal{C}_{\text{gap}}^{\text{mag}} = +1$). This agrees with the tight-binding prediction [Eq. (28)], which establishes that the gap Chern number has absolute value 1 and the same sign as the nonreciprocal hopping $\tilde{\tau}_A = \tilde{\tau}_A^{++}$ between second-neighbors in sublattice $A$, linked by the vector $\gamma_1$. Indeed, numerical estimates of the integrals in Eq. (B12) which determine the hopping yield $\tilde{\tau}_A \cong -5.0 \times 10^{-4}$ ($\tilde{\tau}_A \cong 2.0 \times 10^{-3}$) for the gyromagnetic (magnetoelectric) even-even perturbation. In the case of the gyromagnetic response, the sign of the gap Chern number can be predicted by inspecting the cross product in the corresponding integrand in Eq. (B12), specifically for the leftmost inclusions in the unit cell. These inclusions are the closest to the nodes in sublattice $A$ and thus contribute the most to the hopping.

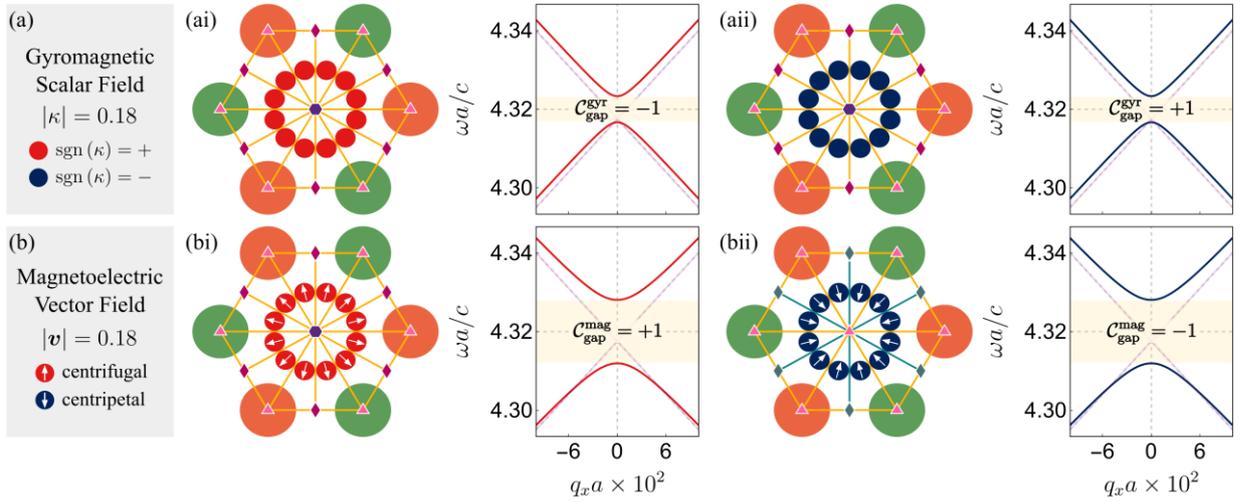

**Fig. 7** - Unit cells, band structures, and gap Chern numbers for graphenes perturbed by nonreciprocal fields with even-even parity. Top row (a): gyromagnetic field with $\kappa_m = 0.18$. Bottom row (b): magnetoelectric field with $v_m = 0.18$. (a*i*, b*i*) Even-even models with gap Chern number $-1$ and $+1$, respectively. (a*ii*, b*ii*) Even-even models obtained from (a*i*, b*i*) via the transformations $\kappa(\mathbf{r}) \to -\kappa(\mathbf{r})$ and $\mathbf{v}(\mathbf{r}) \to -\mathbf{v}(\mathbf{r})$. The gap Chern number is $+1$ and $-1$, respectively. The frequency bands are shown in the vicinity of the Dirac points $K^+$ (solid lines) and $K^-$ (dashed lines), and are represented with the same color as the nonreciprocal elements in the corresponding hexagonal unit cells.

The parity classes of the gyromagnetic fields in Fig. 5(b) are unchanged if we switch the algebraic sign of the response, $\kappa(\mathbf{r}) \to -\kappa(\mathbf{r})$. Likewise, transforming $\mathbf{v}(\mathbf{r}) \to -\mathbf{v}(\mathbf{r})$, i.e., turning the centrifugal magnetoelectric vectors into centripetal, and vice-versa, preserves the parities in Fig. 5(c). For the gapped systems, these transformations preserve parity but switch the

-28-

sign of the gap Chern number. In fact, the gap Chern number has the sign of the nonreciprocal hopping between second neighbors. This hopping is an overlap integral whose integrand is linear with respect to the gyromagnetic/magnetoelectric field. Therefore, a change of sign of the field necessarily affects the gap Chern number. In Figs. 7(a*ii*) and 7(b*ii*), we show the unit cells of the crystals resulting from the transformations $\kappa(\mathbf{r}) \to -\kappa(\mathbf{r})$ and $\mathbf{v}(\mathbf{r}) \to -\mathbf{v}(\mathbf{r})$ of the original crystals in Figs. 7(a*i*) and 7(b*i*). The band structures remain identical, but the gap Chern numbers switch sign: $\mathcal{C}_{\text{gap}}^{\text{gyr}} = +1$ and $\mathcal{C}_{\text{gap}}^{\text{mag}} = -1$. Note that even-even nonreciprocal perturbations only originate a split term. Consequently, the frequency dispersion is insensitive to the sign change of $\kappa$ and $\mathbf{v}$ [see Eq. (22)].

## VIII. Conclusion

In conclusion, we have developed a comprehensive symmetry-based classification of Chern phases in honeycomb photonic crystals, focusing on arbitrary nonreciprocal couplings, including gyromagnetic, pseudo-Tellegen, and moving medium responses. Through a tight-binding model, we demonstrate that while nonreciprocal interactions are crucial, they alone are insufficient to induce a topologically nontrivial phase. Instead, a nontrivial p6m component in the nonreciprocal fields is necessary to open a bandgap and achieve a non-zero Chern number. Our findings provide a symmetry-based roadmap for designing and engineering topological phases in graphene-like photonic crystals, offering clear guidelines for leveraging nonreciprocal perturbations to create nontrivial topological systems.

**Acknowledgements:** This work is supported in part by by the Institution of Engineering and Technology (IET), by the Simons Foundation under the award 733700 (Simons Collaboration in Mathematics and Physics, "Harnessing Universal Symmetry Concepts for Extreme Wave Phenomena"), and by Fundação para a Ciência e a Tecnologia and Instituto de Telecomunicações under project UIDB/50008/2020.

## Appendix A: Tight Binding Model of Photonic Graphene

In the spectral domain, the source-free Maxwell's equations for TE states in photonic graphene can be compactly formulated as an eigenvalue problem:

$$\mathcal{H}_c \psi = \omega \psi. \tag{A1}$$

Here, $\psi = (E_z, H_x Z_0, H_y Z_0)^\mathrm{T}$ is a state vector written in terms of the 3 nonvanishing components of the electromagnetic field, and the crystal Hamiltonian is defined by:

$$\mathcal{H}_c = \mathcal{M}_c^{-1} \cdot \mathcal{D}, \quad \text{with} \tag{A2a}$$

$$\mathcal{M}_c = \begin{pmatrix} \varepsilon_c & 0 & 0 \\ 0 & 1 & 0 \\ 0 & 0 & 1 \end{pmatrix} \text{ and } \mathcal{D} \equiv ic \begin{pmatrix} 0 & -\partial_y & \partial_x \\ -\partial_y & 0 & 0 \\ \partial_x & 0 & 0 \end{pmatrix}. \tag{A2b}$$

We use the symbols $\cdot$ and $\mathrm{T}$ to denote the matrix multiplication and transpose operators, respectively. The material matrix $\mathcal{M}_c$ is defined in terms of the relative permittivity of the crystal, $\varepsilon_c$.

We look for crystal modes which are well-approximated by linear combinations of "atomic" orbitals $\boldsymbol{\varphi}_{\alpha,\mathbf{n}}$ localized around the nodes of the honeycomb lattice. This is the so-called tight binding *ansatz*:

$$\psi(\mathbf{r}) \approx c_A g_A(\mathbf{r}) + c_B g_B(\mathbf{r}), \text{ with} \tag{A3a}$$

$$g_\alpha(\mathbf{r}) \approx \mathcal{N}^{-1/2} \sum_{\mathbf{n}} e^{i\mathbf{k} \cdot (\mathbf{R}_{\alpha,\mathbf{n}} - \mathbf{R}_{\alpha,0})} \boldsymbol{\varphi}_{\alpha,\mathbf{n}}(\mathbf{r}), \tag{A3b}$$

written here in a way which automatically satisfies Bloch's theorem, $\psi(\mathbf{r} + \mathbf{R}_{A,\mathbf{n}}) = e^{i\mathbf{k} \cdot \mathbf{R}_{A,\mathbf{n}}} \psi(\mathbf{r})$. The coefficients $c_A$ and $c_B$ are arbitrary complex numbers and $\mathcal{N}$ is a real-valued normalization factor to be fixed later. The quasimomentum $\mathbf{k}$ is in the FBZ of the crystal. As discussed in Sec. II, the orbital $\boldsymbol{\varphi}_{\alpha,\mathbf{n}}$ corresponds to the ground-state of single-rod system with eigenfrequency $\omega_0$. The subscript $(\alpha, \mathbf{n})$ indicates that the rod is centered at the honeycomb node with coordinates $\mathbf{R}_{\alpha,\mathbf{n}}$ ($\alpha \in \{A, B\}$ and $\mathbf{n} \in \mathbb{Z}^2$). Evidently, $\boldsymbol{\varphi}_{\alpha,\mathbf{n}}$ satisfies $\mathcal{H}_{\alpha,\mathbf{n}} \boldsymbol{\varphi}_{\alpha,\mathbf{n}} = \omega_0 \boldsymbol{\varphi}_{\alpha,\mathbf{n}}$, being $\mathcal{H}_{\alpha,\mathbf{n}}$ defined in the same manner as $\mathcal{H}_c$, with the permittivity of the crystal $\varepsilon_c$ replaced by the permittivity of single rod system $\varepsilon_{\alpha,\mathbf{n}}$. For convenience, we denote the material matrix of the single rod system as $\mathcal{M}_{\alpha,\mathbf{n}}$.

From Eqs. (3) and (4), the atomic orbitals can be explicitly written as

$$\boldsymbol{\varphi}_{\alpha,\mathbf{n}}(\mathbf{r}) = \begin{pmatrix} e_{\alpha,\mathbf{n}}(\mathbf{r}) \\ -ih_{\alpha,\mathbf{n}}(\mathbf{r})(\hat{\mathbf{z}} \times \hat{\mathbf{u}}_{\alpha,\mathbf{n}}) \cdot \hat{\mathbf{x}} \\ -ih_{\alpha,\mathbf{n}}(\mathbf{r})(\hat{\mathbf{z}} \times \hat{\mathbf{u}}_{\alpha,\mathbf{n}}) \cdot \hat{\mathbf{y}} \end{pmatrix}. \tag{A4}$$

We defined the dimensionless field amplitudes $e(\mathbf{r}) = E_{z,0}/E_0$ and $h(\mathbf{r}) = iZ_0 H_{\phi,0}/E_0$ from Eqs. (3) and (4), and use the notation $f_{\alpha,\mathbf{n}}(\mathbf{r}) \equiv f(\mathbf{r} - \mathbf{R}_{\alpha,\mathbf{n}})$ for $f = e, h$. Furthermore, we introduced the unit vector

$$\hat{\mathbf{u}}_{\alpha,\mathbf{n}} = \frac{\mathbf{r} - \mathbf{R}_{\alpha,\mathbf{n}}}{|\mathbf{r} - \mathbf{R}_{\alpha,\mathbf{n}}|}, \tag{A5}$$

which is directed from the node $\mathbf{R}_{\alpha,\mathbf{n}}$ to the point $\mathbf{r}$ of the crystal plane.

In Eq. (A1), the electrodynamics is formulated as a Schrodinger-type equation. Therefore, we can readily extend the standard method of creating tight binding models of quantum matter to our photonic platform. Namely, we deduce the $2 \times 2$ matrix representation $\mathcal{H}_\mathbf{k}$ of the Hamiltonian $\mathcal{H}_c$ on the space of crystal states spanned by the functions $g_A$ and $g_B$ [see Eq. (A3b)]. To do this, we first set a suitable inner product in that space. Because the electrodynamics is conservative in the photonic graphene, the frequency spectrum of the Hamiltonian $\mathcal{H}_c$ must be real-valued. This prompts the choice of an inner-product $\langle \psi | \psi' \rangle$ between two arbitrary crystal eigenstates $\psi$ and $\psi'$ for which $\mathcal{H}_c$ is self-adjoint:

$$\langle \psi | \psi' \rangle \equiv \int_{\text{u.c.}} d^2\mathbf{r}\, \psi^\dagger(\mathbf{r}) \cdot \mathcal{M}_c(\mathbf{r}) \cdot \psi'(\mathbf{r}). \tag{A6}$$

The integration is performed over one unit cell (u.c.) of the honeycomb lattice, as is usual for Bloch-periodic functions. Notably, the crystal states change by a phase factor under a lattice translation, but the inner-product is insensitive to the unit cell choice. It should be noted that $\mathcal{M}_c$ is indefinite in the metal ($\varepsilon < 0$), and so, seemingly, the inner-product does not satisfy the requirement of positive definiteness. However, as the atomic orbitals are strongly localized in the air region, the inner product satisfies $\langle \psi | \psi \rangle > 0$ for any state described by the tight-binding model.

Using Eqs. (A3b) and (A6), we find that

$$\langle g_\alpha | g_\beta \rangle \approx \delta_{\alpha,\beta} \, \mathcal{N}^{-1} \sum_{\mathbf{n}} \int_{\text{u.c.}} d^2\mathbf{r} \, \boldsymbol{\varphi}_{\alpha,\mathbf{n}}(\mathbf{r})^\dagger \cdot \mathcal{M}_c(\mathbf{r}) \cdot \boldsymbol{\varphi}_{\alpha,\mathbf{n}}(\mathbf{r}). \tag{A7}$$

We used the approximation, $\boldsymbol{\varphi}^\dagger_{\alpha,\mathbf{n}}(\mathbf{r}) \cdot \mathcal{M}_c(\mathbf{r}) \cdot \boldsymbol{\varphi}_{\beta,\mathbf{m}}(\mathbf{r}) \approx 0$, which holds true in the tight-binding picture when the two lattice nodes are different. The result above can be written more simply as

$$\langle g_\alpha | g_\beta \rangle \approx \delta_{\alpha,\beta} \, \mathcal{N}^{-1} \left( \boldsymbol{\varphi}_{\alpha,\mathbf{0}} | \mathcal{M}_c \cdot \boldsymbol{\varphi}_{\alpha,\mathbf{0}} \right), \tag{A8}$$

where

$$\left( \boldsymbol{\varphi}_{\alpha,\mathbf{n}} | \boldsymbol{\varphi}_{\beta,\mathbf{m}} \right) \equiv \int_{\mathbb{R}^2} d^2\mathbf{r} \, \boldsymbol{\varphi}_{\alpha,\mathbf{n}}(\mathbf{r})^\dagger \cdot \boldsymbol{\varphi}_{\beta,\mathbf{m}}(\mathbf{r}) \tag{A9}$$

is the canonical inner-product in the space of atomic orbitals. Here, the integration is over the entire crystal plane, $\mathbb{R}^2$, rather than over a single unit cell. We took into account that all translations of the initial unit cell combined tile the crystal plane. We fix the normalization factor as

$$\mathcal{N} = \left( \boldsymbol{\varphi}_{A,\mathbf{0}} | \mathcal{M}_c \cdot \boldsymbol{\varphi}_{A,\mathbf{0}} \right). \tag{A10}$$

For this choice, the tight-binding basis is orthonormalized, $\langle g_\alpha | g_\beta \rangle \approx \delta_{\alpha,\beta}$ (see Eq. (A8) and note that $\left( \boldsymbol{\varphi}_{\alpha,\mathbf{0}} | \mathcal{M}_c \cdot \boldsymbol{\varphi}_{\alpha,\mathbf{0}} \right)$ is independent of the sublattice). Then, the $2 \times 2$ matrix representation $\mathcal{H}_\mathbf{k} = \left[ h_{\alpha\beta} \right]$ of the Hamiltonian $\mathcal{H}_c$ in this basis has elements

$$h_{\alpha\beta} = \langle g_\alpha | \mathcal{H}_c g_\beta \rangle. \tag{A11}$$

We dropped the subscript $\mathbf{k}$ for conciseness.

It is useful to note that the Hamiltonians of the crystal and single-rod system are related as

$$\mathcal{H}_c = \mathcal{M}_c^{-1} \cdot \mathcal{M}_{\alpha,\mathbf{n}} \cdot \mathcal{H}_{\alpha,\mathbf{n}}. \tag{A12}$$

Feeding this identity into Eq. (A11) yields

$$h_{\alpha\beta} = \omega_0 \, \mathcal{N}^{-1} e^{i\mathbf{k}\cdot(\mathbf{R}_{\alpha,0}-\mathbf{R}_{\beta,0})} \sum_{\mathbf{n},\mathbf{m}} e^{i\mathbf{k}\cdot(\mathbf{R}_{\beta,\mathbf{m}}-\mathbf{R}_{\alpha,\mathbf{n}})} \int_{\text{u.c.}} d^2\mathbf{r}\, \boldsymbol{\varphi}_{\alpha,\mathbf{n}}(\mathbf{r})^\dagger \cdot \mathcal{M}_{\beta,\mathbf{m}} \cdot \boldsymbol{\varphi}_{\beta,\mathbf{m}}(\mathbf{r}). \tag{A13}$$

We can rewrite this result as

$$h_{\alpha\beta} = \omega_0 \, \mathcal{N}^{-1} e^{-i\mathbf{k}\cdot\mathbf{R}_{\beta,0}} \sum_{\mathbf{n},\mathbf{m}} e^{i\mathbf{k}\cdot\mathbf{R}_{\beta,\mathbf{m}-\mathbf{n}}} \int_{\text{u.c.}} d^2\mathbf{r}\, \boldsymbol{\varphi}_{\alpha,\mathbf{n}}(\mathbf{r})^\dagger \cdot \mathcal{M}_{\beta,\mathbf{m}} \cdot \boldsymbol{\varphi}_{\beta,\mathbf{m}}(\mathbf{r}). \tag{A14}$$

where we used $\mathbf{R}_{\beta,\mathbf{m}} - \mathbf{R}_{\alpha,\mathbf{n}} = \mathbf{R}_{\beta,\mathbf{m}-\mathbf{n}} - \mathbf{R}_{\alpha,0}$. Next, we replace $\mathbf{m}-\mathbf{n} \to \mathbf{m}'$ in the double sum and note that $\int_{\text{u.c.}} d^2\mathbf{r}\, \boldsymbol{\varphi}_{\alpha,\mathbf{n}}(\mathbf{r})^\dagger \cdot \mathcal{M}_{\beta,\mathbf{m}'+\mathbf{n}} \cdot \boldsymbol{\varphi}_{\beta,\mathbf{m}'+\mathbf{n}}(\mathbf{r}) = \int_{\text{u.c.}-\mathbf{R}_{\alpha,\mathbf{n}}+\mathbf{R}_{\alpha,0}} d^2\mathbf{r}\, \boldsymbol{\varphi}_{\alpha,0}(\mathbf{r})^\dagger \cdot \mathcal{M}_{\beta,\mathbf{m}'} \cdot \boldsymbol{\varphi}_{\beta,\mathbf{m}'}(\mathbf{r})$. Evidently, the outstanding sum over $\mathbf{n}$ tiles the plane. This means that $h_{\alpha\beta}$ is determined by orbital inner-products as

$$h_{\alpha\beta} = \omega_0 \, \mathcal{N}^{-1} e^{-i\mathbf{k}\cdot\mathbf{R}_{\beta,0}} \sum_{\mathbf{m}} e^{i\mathbf{k}\cdot\mathbf{R}_{\beta,\mathbf{m}}} \left(\boldsymbol{\varphi}_{\alpha,0} \,|\, \mathcal{M}_{\beta,\mathbf{m}} \cdot \boldsymbol{\varphi}_{\beta,\mathbf{m}}\right). \tag{A15}$$

The integrals $\left(\boldsymbol{\varphi}_{\alpha,0} \,|\, \mathcal{M}_{\beta,\mathbf{m}} \cdot \boldsymbol{\varphi}_{\beta,\mathbf{m}}\right)$ above measure the orbital overlaps mediated by the electromagnetic response of the single-rod system. Due to the dispersion of the metal permittivity in the host region, these overlaps depend on the frequency $\omega$, and, thereby, so do the matrix elements $h_{\alpha\beta} = h_{\alpha\beta}(\omega)$. For simplicity, we approximate $h_{\alpha\beta}(\omega) \approx h_{\alpha\beta}(\omega_0)$. This simplification allows us to calculate the photonic crystal dispersion $\omega$ vs $\mathbf{k}$ via standard diagonalization of the tight-binding Hamiltonian $\mathcal{H}_\mathbf{k}(\omega_0) = \left[h_{\alpha\beta}(\omega_0)\right]$. This approximation is justified in the tight-binding limit, where the orbital overlaps are small, and the eigenfrequencies $\omega$ remain close to $\omega_0$ throughout the entire FBZ. Note that the most relevant dispersion effects are already accounted for in the modeling of the single-rod system.

Next, we write $\mathcal{M}_{\beta,\mathbf{m}}(\omega_0) = \mathcal{M}_c(\omega_0) - \Delta_{\beta,\mathbf{m}}(\omega_0)$, with

$$\Delta_{\beta,\mathbf{m}} \equiv \sum_{\mathbf{R}_{\alpha,\mathbf{n}}}^{\neq \mathbf{R}_{\beta,\mathbf{m}}} \frac{\omega_p^2}{\omega_0^2} \Theta\left(R - \left|\mathbf{r} - \mathbf{R}_{\alpha,\mathbf{n}}\right|\right) \begin{pmatrix} 1 & 0 & 0 \\ 0 & 0 & 0 \\ 0 & 0 & 0 \end{pmatrix} \tag{A16}$$

the difference between the responses of the crystal and the single-rod system. The sum is over all nodes of the honeycomb grid except the one labeled by the index $(\beta,\mathbf{m})$. Accordingly, the orbital overlaps split as

$$\left(\boldsymbol{\varphi}_{\alpha,\mathbf{0}} \mid \mathcal{M}_{\beta,\mathbf{m}} \cdot \boldsymbol{\varphi}_{\beta,\mathbf{m}}\right) = \left(\boldsymbol{\varphi}_{\alpha,\mathbf{0}} \mid \mathcal{M}_c \cdot \boldsymbol{\varphi}_{\beta,\mathbf{m}}\right) - \left(\boldsymbol{\varphi}_{\alpha,\mathbf{0}} \mid \Delta_{\beta,\mathbf{m}} \cdot \boldsymbol{\varphi}_{\beta,\mathbf{m}}\right). \tag{A17}$$

When the orbitals $\boldsymbol{\varphi}_{\alpha,\mathbf{0}}$ and $\boldsymbol{\varphi}_{\beta,\mathbf{m}}$ are centered at different lattice nodes and are strongly localized in the air region, the overlap $\left(\boldsymbol{\varphi}_{\alpha,\mathbf{0}} \mid \Delta_{\beta,\mathbf{m}} \cdot \boldsymbol{\varphi}_{\beta,\mathbf{m}}\right)$ dominates $\left(\boldsymbol{\varphi}_{\alpha,\mathbf{0}} \mid \mathcal{M}_c \cdot \boldsymbol{\varphi}_{\beta,\mathbf{m}}\right)$. This is true because $\omega_p^2 / \omega_0^2 \gg 1$ in the tight-binding regime. Thus, in the calculation of $h_{\alpha\beta}$, we can apply the replacement $\left(\boldsymbol{\varphi}_{\alpha,\mathbf{0}} \mid \mathcal{M}_c \cdot \boldsymbol{\varphi}_{\beta,\mathbf{m}}\right) \to \delta_{\alpha,\beta}\delta_{\mathbf{0},\mathbf{m}} \left(\boldsymbol{\varphi}_{\alpha,\mathbf{0}} \mid \mathcal{M}_c \cdot \boldsymbol{\varphi}_{\alpha,\mathbf{0}}\right)$, which leads to:

$$h_{\alpha\beta} \approx \omega_0 \left( \delta_{\alpha,\beta} - \mathcal{N}^{-1} e^{-i\mathbf{k}\cdot\mathbf{R}_{\beta,0}} \sum_{\mathbf{m}} e^{i\mathbf{k}\cdot\mathbf{R}_{\beta,\mathbf{m}}} \left(\boldsymbol{\varphi}_{\alpha,\mathbf{0}} \mid \Delta_{\beta,\mathbf{m}} \cdot \boldsymbol{\varphi}_{\beta,\mathbf{m}}\right) \right), \tag{A18}$$

For simplicity, we employed the additional approximation $\left(\boldsymbol{\varphi}_{\alpha,\mathbf{0}} \mid \mathcal{M}_c \cdot \boldsymbol{\varphi}_{\alpha,\mathbf{0}}\right) \approx \mathcal{N}$, justified by the fact that $\mathcal{M}_c$ is the identity matrix at the relevant lattice site.

Next, consider interactions only up to next nearest neighbors. We build a dimensionless tight-binding Hamiltonian $\mathcal{H}_\mathbf{k} \to \mathcal{H}_\mathbf{k}/\omega_0 = [h_{\alpha\beta}/\omega_0]$ that reads

$$\mathcal{H}_\mathbf{k} \approx \mathbf{1}_{2\times 2} - \Delta_\mathbf{k}, \tag{A19}$$

where

$$\Delta_\mathbf{k} = \begin{pmatrix} \Delta_{A,0} + \sum_i \Delta_{AA}(\boldsymbol{\gamma}_i) e^{i\mathbf{k}\cdot\boldsymbol{\gamma}_i} & e^{i\mathbf{k}\cdot\boldsymbol{\delta}_2} \sum_i \Delta_{AB}(-\boldsymbol{\delta}_i) e^{-i\mathbf{k}\cdot\boldsymbol{\delta}_i} \\ e^{-i\mathbf{k}\cdot\boldsymbol{\delta}_2} \sum_i \Delta_{BA}(\boldsymbol{\delta}_i) e^{i\mathbf{k}\cdot\boldsymbol{\delta}_i} & \Delta_{B,0} + \sum_i \Delta_{BB}(\boldsymbol{\gamma}_i) e^{i\mathbf{k}\cdot\boldsymbol{\gamma}_i} \end{pmatrix}, \tag{A20}$$

with the sums extended only to the relevant neighbors and

$$\begin{aligned}
\Delta_{\alpha,0} &= \mathcal{N}^{-1}\left(\varphi_{\mathbf{R}_{\alpha,0}} \mid \Delta_{\mathbf{R}_{\alpha,0}} \cdot \varphi_{\mathbf{R}_{\alpha,0}}\right) \\
\Delta_{\alpha\alpha}(\boldsymbol{\gamma}_i) &= \mathcal{N}^{-1}\left(\varphi_{\mathbf{R}_{\alpha,0}} \mid \Delta_{\mathbf{R}_{\alpha,0}+\boldsymbol{\gamma}_i} \cdot \varphi_{\mathbf{R}_{\alpha,0}+\boldsymbol{\gamma}_i}\right) \\
\Delta_{AB}(-\boldsymbol{\delta}_i) &= \mathcal{N}^{-1}\left(\varphi_{\mathbf{R}_{A,0}} \mid \Delta_{\mathbf{R}_{A,0}-\boldsymbol{\delta}_i} \cdot \varphi_{\mathbf{R}_{A,0}-\boldsymbol{\delta}_i}\right) \\
\Delta_{BA}(\boldsymbol{\delta}_i) &= \mathcal{N}^{-1}\left(\varphi_{\mathbf{R}_{B,0}} \mid \Delta_{\mathbf{R}_{B,0}+\boldsymbol{\delta}_i} \cdot \varphi_{\mathbf{R}_{B,0}+\boldsymbol{\delta}_i}\right)
\end{aligned} \quad (A21)$$

For convenience, we set $\varphi_{\mathbf{R}_{\alpha,\mathbf{n}}} = \varphi_{\alpha,\mathbf{n}}$ and $\Delta_{\mathbf{R}_{\alpha,\mathbf{n}}} = \Delta_{\alpha,\mathbf{n}}$, i.e., we refer to a lattice node via its position vector. Below, this notation is extended to other quantities. The vectors $\boldsymbol{\delta}_i$ ($\boldsymbol{\gamma}_i$) join first (second) neighbors [see Fig. 3(a) of the main text]. The eigenvalues of the tight-binding Hamiltonian in Eq. (A19) determine the band structure of the photonic graphene:

$$\det\left(\mathcal{H}_{\mathbf{k}} - \frac{\omega}{\omega_0}\mathbf{1}_{2\times 2}\right) = 0. \quad (A22)$$

In remainder of this Appendix, we derive the integral expressions for the tight-binding parameters presented in Eq. (7) of the main text. Using Eqs. (A4) and (A9), we can write an orbital overlap from $\Delta_{\mathbf{k}}$ as

$$\left(\varphi_{\mathbf{R}_{\alpha,0}} \mid \Delta_{\mathbf{R}_{\beta,\mathbf{m}}} \cdot \varphi_{\mathbf{R}_{\beta,\mathbf{m}}}\right) = \sum_{\mathbf{R}_{\delta,\mathbf{p}}}^{\neq \mathbf{R}_{\beta,\mathbf{m}}} \frac{\omega_p^2}{\omega_0^2} \int_{D_{\delta,\mathbf{p}}} d^2\mathbf{r}\, e_{\mathbf{R}_{\alpha,0}} e_{\mathbf{R}_{\beta,\mathbf{m}}}. \quad (A23)$$

Here $D_{\delta,\mathbf{p}}$ represents the circle on the $xoy$-plane with radius $R$ and centered at $\mathbf{R}_{\delta,\mathbf{p}}$. Because $e_{\mathbf{R}_{\alpha,\mathbf{n}}}(\mathbf{r})$ only depends on the distance from $\mathbf{r}$ to $\mathbf{R}_{\alpha,\mathbf{n}}$, it is straightforward to show that $\Delta_{A,0} = \Delta_{B,0}$. Similarly, $\Delta_{\alpha\alpha}(\boldsymbol{\gamma}_i)$ and $\Delta_{BA}(\boldsymbol{\delta}_i)$ are independent of the considered neighbor ($i$). Furthermore, $\Delta_{AA}(\boldsymbol{\gamma}_i) = \Delta_{BB}(\boldsymbol{\gamma}_i)$ and $\Delta_{AB}(-\boldsymbol{\delta}_i) = \Delta_{BA}(\boldsymbol{\delta}_i)$. These properties considerably simplify the matrix $\Delta_{\mathbf{k}}$ in Eq. (A20). We ultimately obtain the three tight-binding parameters shown in Eq. (7) of the main text: the self-frequency $\Omega = \Delta_{A,0}$, the hopping $t = \Delta_{BA}(\boldsymbol{\delta}_1)$ between first neighbors, and the hopping $\tilde{t} = \Delta_{AA}(\boldsymbol{\gamma}_1)$ between second neighbors. Their integral-form expressions follow from the general one in Eq. (A23);

$$\Omega = \mathcal{N}^{-1} \sum_{\mathbf{R}_{\alpha,\mathbf{n}}}^{\neq \mathbf{R}_{A,0}} \frac{\omega_p^2}{\omega_0^2} \int_{D_{\alpha,\mathbf{n}}} d^2\mathbf{r}\, e_{\mathbf{R}_{A,0}}^2,$$

$$t = \mathcal{N}^{-1} \sum_{\mathbf{R}_{\alpha,\mathbf{n}}}^{\neq \mathbf{R}_{B,0}+\boldsymbol{\delta}_1} \frac{\omega_p^2}{\omega_0^2} \int_{D_{\alpha,\mathbf{n}}} d^2\mathbf{r}\, e_{\mathbf{R}_{B,0}} e_{\mathbf{R}_{B,0}+\boldsymbol{\delta}_1}, \qquad (A24)$$

$$\tilde{t} = \mathcal{N}^{-1} \sum_{\mathbf{R}_{\alpha,\mathbf{n}}}^{\neq \mathbf{R}_{A,0}+\boldsymbol{\gamma}_1} \frac{\omega_p^2}{\omega_0^2} \int_{D_{\alpha,\mathbf{n}}} d^2\mathbf{r}\, e_{\mathbf{R}_{A,0}} e_{\mathbf{R}_{A,0}+\boldsymbol{\gamma}_1}.$$

**Appendix B: Tight Binding Model of Modified Graphene**

As discussed in the main text, we suppose that the response of the "pristine" photonic graphene is modified with a nonreciprocal part that preserves the decoupling of TE modes from other excitations. In this case, the Hamiltonian takes the form

$$\hat{\mathcal{H}}'_c = (\mathcal{M}_c + \mathcal{P})^{-1} \cdot \mathcal{D}, \qquad (B1)$$

where $\mathcal{P} = \mathcal{K} + \mathcal{V}$ is written in terms of the representations of the gyromagnetic and magnetoelectric dyadics in the space of TE-polarized states

$$\mathcal{K} = i\kappa \begin{pmatrix} 0 & 0 & 0 \\ 0 & 0 & 1 \\ 0 & -1 & 0 \end{pmatrix} \quad \text{and} \quad \mathcal{V} = \begin{pmatrix} 0 & v_x & v_y \\ v_x & 0 & 0 \\ v_y & 0 & 0 \end{pmatrix}, \qquad (B2)$$

respectively. We use the prime symbol to distinguish the Hamiltonian for the nonreciprocal crystal from the Hamiltonian of the bare photonic graphene.

To develop the tight-binding model of the perturbed crystal, we follow the same steps as in Appendix A. Because the gyromagnetic and magnetoelectric responses are conservative, the frequency spectrum of the Hamiltonian of the perturbed graphene $\hat{\mathcal{H}}'_c$ remains real-valued. Analogous to Eq. (A12), it is possible to relate the Hamiltonian of the perturbed crystal with the Hamiltonian of a single rod system $\hat{\mathcal{H}}'_c = (\mathcal{M}_c + \mathcal{P})^{-1} \cdot \mathcal{M}_{\alpha,\mathbf{n}} \cdot \mathcal{H}_{\alpha,\mathbf{n}}$. For a weak nonreciprocal perturbation, we can use $(\mathbf{1}_{3\times 3} + \mathcal{P} \cdot \mathcal{M}_c^{-1})^{-1} \approx \mathbf{1}_{3\times 3} - \mathcal{P} \cdot \mathcal{M}_c^{-1}$, so that

$$\hat{\mathcal{H}}'_c \approx \mathcal{M}_c^{-1} \cdot (\mathcal{M}_{\alpha,\mathbf{n}} - \mathcal{P} + \mathcal{P} \cdot \Delta_{\alpha,\mathbf{n}}) \cdot \mathcal{H}_{\alpha,\mathbf{n}}. \qquad (B3)$$

Using this identity, we write the element $h'_{\alpha\beta} = \langle g_\alpha | \hat{\mathcal{H}}'_c \, g_\beta \rangle$ of the tight-binding Hamiltonian $\mathcal{H}'_\mathbf{k} = [h'_{\alpha\beta}]$ for the modified graphene as

$$h'_{\alpha\beta} \approx h_{\alpha\beta} - \omega_0 \, \mathcal{N}^{-1} e^{-i\mathbf{k}\cdot\mathbf{R}_{\beta,0}} \sum_\mathbf{m} e^{i\mathbf{k}\cdot\mathbf{R}_{\beta,\mathbf{m}}} \left( (\boldsymbol{\varphi}_{\alpha,\mathbf{0}} | \mathcal{P} \cdot \boldsymbol{\varphi}_{\beta,\mathbf{m}}) - (\boldsymbol{\varphi}_{\alpha,\mathbf{0}} | \mathcal{P} \cdot \Delta_{\beta,\mathbf{m}} \cdot \boldsymbol{\varphi}_{\beta,\mathbf{m}}) \right). \tag{B4}$$

Here, $h_{\alpha\beta}$, $\mathcal{N}$, and the inner product are defined in Appendix A. As before, we disregard all interactions between third or higher-order neighbors.

The terms $(\boldsymbol{\varphi}_{\alpha,\mathbf{0}} | \mathcal{P} \cdot \Delta_{\beta,\mathbf{m}} \cdot \boldsymbol{\varphi}_{\beta,\mathbf{m}})$ above can be dropped. This can be justified as follows. For the gyromagnetic class, the product vanishes identically, $\mathcal{K} \cdot \Delta_{\beta,\mathbf{m}} = \mathbf{0}_{3\times 3}$ [see Eqs. (B2) and (A15)]. On the contrary, for the magnetoelectric responses, the product is generally nonvanishing inside the crystal air discs ($D_{\alpha,\mathbf{n}}$). However, provided the magnetoelectric coupling is dominantly placed outside the air discs, one can safely neglect the overlap $(\boldsymbol{\varphi}_{\alpha,\mathbf{0}} | \mathcal{V} \cdot \Delta_{\beta,\mathbf{m}} \cdot \boldsymbol{\varphi}_{\beta,\mathbf{m}})$ as compared to $(\boldsymbol{\varphi}_{\alpha,\mathbf{0}} | \mathcal{V} \cdot \boldsymbol{\varphi}_{\beta,\mathbf{m}})$ which has the entire crystal plane as its domain of integration. As a result, Eq. (B4) becomes

$$h'_{\alpha\beta} \approx h_{\alpha\beta} - \omega_0 \, \mathcal{N}^{-1} e^{-i\mathbf{k}\cdot\mathbf{R}_{\beta,0}} \sum_\mathbf{m} e^{i\mathbf{k}\cdot\mathbf{R}_{\beta,\mathbf{m}}} (\boldsymbol{\varphi}_{\alpha,\mathbf{0}} | \mathcal{P} \cdot \boldsymbol{\varphi}_{\beta,\mathbf{m}}). \tag{B5}$$

Hence, we can write the dimensionless tight-binding Hamiltonian $\mathcal{H}'_\mathbf{k} \to \mathcal{H}'_\mathbf{k}/\omega_0 = [h'_{\alpha\beta}/\omega_0]$ of the perturbed crystal as

$$\mathcal{H}'_\mathbf{k} \approx \mathcal{H}_\mathbf{k} - \mathcal{K}_\mathbf{k} - \mathcal{V}_\mathbf{k}, \tag{B6}$$

where the last two matrices are obtained from replacing $\mathcal{O} \to \mathcal{K}$ and $\mathcal{O} \to \mathcal{V}$ in the template

$$\mathcal{O}_\mathbf{k} = \begin{pmatrix} \mathcal{O}_{A,0} + \sum_i \mathcal{O}_{AA}(\boldsymbol{\gamma}_i) e^{i\mathbf{k}\cdot\boldsymbol{\gamma}_i} & e^{i\mathbf{k}\cdot\boldsymbol{\delta}_2} \sum_i \mathcal{O}_{AB}(-\boldsymbol{\delta}_i) e^{-i\mathbf{k}\cdot\boldsymbol{\delta}_i} \\ e^{-i\mathbf{k}\cdot\boldsymbol{\delta}_2} \sum_i \mathcal{O}_{BA}(\boldsymbol{\delta}_i) e^{i\mathbf{k}\cdot\boldsymbol{\delta}_i} & \mathcal{O}_{B,0} + \sum_i \mathcal{O}_{BB}(\boldsymbol{\gamma}_i) e^{i\mathbf{k}\cdot\boldsymbol{\gamma}_i} \end{pmatrix} \tag{B7}$$

with

$$\begin{aligned}
\mathcal{O}_{\alpha,0} &= \mathcal{N}^{-1}\left(\varphi_{\mathbf{R}_{\alpha,0}} \mid \mathcal{O}\cdot\varphi_{\mathbf{R}_{\alpha,0}}\right), \\
\mathcal{O}_{\alpha\alpha}(\boldsymbol{\gamma}_i) &= \mathcal{N}^{-1}\left(\varphi_{\mathbf{R}_{\alpha,0}} \mid \mathcal{O}\cdot\varphi_{\mathbf{R}_{\alpha,0}+\boldsymbol{\gamma}_i}\right), \\
\mathcal{O}_{AB}(-\boldsymbol{\delta}_i) &= \mathcal{N}^{-1}\left(\varphi_{\mathbf{R}_{A,0}} \mid \mathcal{O}\cdot\varphi_{\mathbf{R}_{A,0}-\boldsymbol{\delta}_i}\right), \\
\mathcal{O}_{BA}(\boldsymbol{\delta}_i) &= \mathcal{N}^{-1}\left(\varphi_{\mathbf{R}_{B,0}} \mid \mathcal{O}\cdot\varphi_{\mathbf{R}_{B,0}+\boldsymbol{\delta}_i}\right).
\end{aligned} \tag{B8}$$

The overlaps in Eq. (B8) can be written in integral-form using Eqs. (A4), (A9), and (B2). Namely,

$$\left(\varphi_1 \mid \mathcal{K}\cdot\varphi_2\right) = i\int_{\mathbb{R}^2} d^2\mathbf{r}\,\kappa\,h_1 h_2 \left(\hat{\mathbf{u}}_1 \times \hat{\mathbf{u}}_2\right)\cdot\hat{\mathbf{z}} \tag{B9}$$

and

$$\left(\varphi_1 \mid \mathcal{V}\cdot\varphi_2\right) = i\int_{\mathbb{R}^2} d^2\mathbf{r}\,\left[\left(e_2 h_1 \hat{\mathbf{u}}_1 - e_1 h_2 \hat{\mathbf{u}}_2\right)\times\mathbf{v}\right]\cdot\hat{\mathbf{z}}, \tag{B10}$$

where $\varphi_i$ is an atomic orbital centered at a node of the honeycomb grid which we now label with $i$ for simplicity. The same notation is used to label the electromagnetic fields and the unit vectors defined in Eq. (A5). The integrands above involve the gyromagnetic and magnetoelectric fields, $\kappa = \kappa(\mathbf{r})$ and $\mathbf{v} = (v_x, v_y) = \mathbf{v}(\mathbf{r})$ [see Eq. (B2)].

As explained in the main text, the nonreciprocal fields are assumed symmetric under 3-fold rotations about any node of the honeycomb lattice. As a result, many overlap integrals are not independent. Let us use the symbol $\mathcal{O}$ from the template in Eq. (B7) to simultaneously refer to the gyromagnetic and magnetoelectric contributions. Due to rotational symmetry, the overlaps $\mathcal{O}_{AB}(-\boldsymbol{\delta}_i)$ and $\mathcal{O}_{BA}(\boldsymbol{\delta}_i)$ are independent of the relative position between first neighbors, i.e., are independent of $i \in \{1,2,3\}$. On the other hand, the second neighbor vectors $\boldsymbol{\gamma}_i$ can be split into two sets $i \in \{1,3,5\}$ and $i \in \{2,4,6\}$ which are closed under 3-fold rotations. For each subset, the overlap $\mathcal{O}_{\alpha\alpha}(\boldsymbol{\gamma}_i)$ is also independent of the relative position between neighbors. Yet another important simplification originates from the observation that overlaps with interchanged orbitals $1 \leftrightarrow 2$ differ by a minus sign: $(\varphi_1 \mid \mathcal{O}\cdot\varphi_2) = -(\varphi_2 \mid \mathcal{O}\cdot\varphi_1)$ [see Eqs. (B9) and (B10)]. This property is the hallmark of nonreciprocity in the tight-binding picture. It tells us that the self-frequencies vanish, $\mathcal{O}_{A,0} = \mathcal{O}_{B,0} = 0$. Besides, $\mathcal{O}_{\alpha\alpha}(\gamma_{1,3,5}) = -\mathcal{O}_{\alpha\alpha}(\gamma_{4,6,2})$ because $\gamma_{1,3,5} = -\gamma_{4,6,2}$ and $\mathcal{O}_{AB}(-\boldsymbol{\delta}_i) = -\mathcal{O}_{BA}(\boldsymbol{\delta}_i)$ [see Fig. 3(a) in the main text]. Different from pristine graphene, the second neighbor overlaps in sublattice $A$ are independent from those in sublattice $B$.

All in all, we obtain six different tight-binding parameters, three per each nonreciprocal response type. We refer to them as nonreciprocal hoppings. Two between first neighbors, $\tau_\kappa \equiv -i\mathcal{K}_{BA}(\boldsymbol{\delta}_1)$ and $\tau_\mathbf{v} \equiv -i\mathcal{V}_{BA}(\boldsymbol{\delta}_1)$, and four between second neighbors, $\tilde{\tau}_{\alpha,\kappa} \equiv -i\mathcal{K}_{\alpha\alpha}(\boldsymbol{\gamma}_1)$ and $\tilde{\tau}_{\alpha,\mathbf{v}} \equiv -i\mathcal{V}_{\alpha\alpha}(\boldsymbol{\gamma}_1)$ with $\alpha \in \{A, B\}$. Combining the two nonreciprocal couplings,

$$\begin{aligned} \tau &= \tau_\kappa + \tau_\mathbf{v} \\ \tilde{\tau}_A &= \tilde{\tau}_{A,\kappa} + \tilde{\tau}_{A,\mathbf{v}} \\ \tilde{\tau}_B &= \tilde{\tau}_{B,\kappa} + \tilde{\tau}_{B,\mathbf{v}} \end{aligned} \tag{B11}$$

we get the components of the correction factor $\mathcal{P}_\mathbf{k} = \mathcal{K}_\mathbf{k} + \mathcal{V}_\mathbf{k}$ shown in Eq. (19), with the shift and split terms, $s$ and $m$, given in Eq. (21). The integral-form expressions of the nonreciprocal hoppings are

$$\begin{aligned} \tau_\kappa &= \mathcal{N}^{-1} \int_{\mathbb{R}^2} d^2\mathbf{r}\, \kappa h_{\mathbf{R}_{B,0}} h_{\mathbf{R}_{B,0}+\boldsymbol{\delta}_1} \left(\hat{\mathbf{u}}_{\mathbf{R}_{B,0}} \times \hat{\mathbf{u}}_{\mathbf{R}_{B,0}+\boldsymbol{\delta}_1}\right) \cdot \hat{\mathbf{z}}, \\ \tilde{\tau}_{\alpha,\kappa} &= \mathcal{N}^{-1} \int_{\mathbb{R}^2} d^2\mathbf{r}\, \kappa\, h_{\mathbf{R}_{\alpha,0}} h_{\mathbf{R}_{\alpha,0}+\boldsymbol{\gamma}_1} \left(\hat{\mathbf{u}}_{\mathbf{R}_{\alpha,0}} \times \hat{\mathbf{u}}_{\mathbf{R}_{\alpha,0}+\boldsymbol{\gamma}_1}\right) \cdot \hat{\mathbf{z}}, \\ \tau_\mathbf{v} &= \mathcal{N}^{-1} \int_{\mathbb{R}^2} d^2\mathbf{r} \left[\left(e_{\mathbf{R}_{B,0}+\boldsymbol{\delta}_1} h_{\mathbf{R}_{B,0}} \hat{\mathbf{u}}_{\mathbf{R}_{B,0}} - e_{\mathbf{R}_{B,0}} h_{\mathbf{R}_{B,0}+\boldsymbol{\delta}_1} \hat{\mathbf{u}}_{\mathbf{R}_{B,0}+\boldsymbol{\delta}_1}\right) \times \mathbf{v}\right] \cdot \hat{\mathbf{z}}, \\ \tilde{\tau}_{\alpha,\mathbf{v}} &= \mathcal{N}^{-1} \int_{\mathbb{R}^2} d^2\mathbf{r} \left[\left(e_{\mathbf{R}_{\alpha,0}+\boldsymbol{\gamma}_1} h_{\mathbf{R}_{\alpha,0}} \hat{\mathbf{u}}_{\mathbf{R}_{\alpha,0}} - e_{\mathbf{R}_{\alpha,0}} h_{\mathbf{R}_{\alpha,0}+\boldsymbol{\gamma}_1} \hat{\mathbf{u}}_{\mathbf{R}_{\alpha,0}+\boldsymbol{\gamma}_1}\right) \times \mathbf{v}\right] \cdot \hat{\mathbf{z}}. \end{aligned} \tag{B12}$$

To conclude, we note that our model only captures the nonreciprocal contributions which are linear on the gyromagnetic and magnetoelectric fields.

## Appendix C: Symmetry Constraints on the Nonreciprocal Hoppings

Here, we derive the symmetry-based constraints shown in Eqs. (25) and (26) of the main text for the gyromagnetic component of the hoppings. The analysis can be readily extended for the magnetoelectric part.

To do this, we investigate how the hopping integrals are influenced by mirror transformations. To begin with, we consider the transformation $\mathbf{r} \to \mathbf{r}' = M_x \cdot \mathbf{r}$, i.e., $(x, y) \to (x, -y)$. Figure 8(a) shows that

$$\hat{\mathbf{u}}_{\mathbf{R}_{\alpha,0}}(\mathbf{r}) = M_x \cdot \hat{\mathbf{u}}_{\mathbf{R}_{\alpha,0}+\gamma_1}(\mathbf{r}'),$$
$$h_{\mathbf{R}_{\alpha,0}}(\mathbf{r}) = h_{\mathbf{R}_{\alpha,0}+\gamma_1}(\mathbf{r}'). \tag{C1}$$

As a result, the hopping integral $\tilde{\tau}_\alpha$ [Eq. (B12)] can be written as:

$$\begin{aligned}\tilde{\tau}_{\alpha,\kappa} &= \mathcal{N}^{-1} \int_{\mathbb{R}^2} d^2\mathbf{r}' \, \kappa(M_x \cdot \mathbf{r}') \, h_{\mathbf{R}_{\alpha,0}+\gamma_1} h_{\mathbf{R}_{\alpha,0}} \left[ \left(M_x \cdot \hat{\mathbf{u}}_{\mathbf{R}_{\alpha,0}+\gamma_1}\right) \times \left(M_x \cdot \hat{\mathbf{u}}_{\mathbf{R}_{\alpha,0}}\right) \right] \cdot \hat{\mathbf{z}} \\ &= \mathcal{N}^{-1} \int_{\mathbb{R}^2} d^2\mathbf{r}' \, \kappa(M_x \cdot \mathbf{r}') \, h_{\mathbf{R}_{\alpha,0}} h_{\mathbf{R}_{\alpha,0}+\gamma_1} \left(\hat{\mathbf{u}}_{\mathbf{R}_{\alpha,0}} \times \hat{\mathbf{u}}_{\mathbf{R}_{\alpha,0}+\gamma_1}\right) \cdot \hat{\mathbf{z}}.\end{aligned} \tag{C2}$$

In particular, when the gyromagnetic field exhibits odd symmetry with respect to the mirror $M_x$, such that $\kappa(M_x \cdot \mathbf{r}) = -\kappa(\mathbf{r})$, the above relation implies that $\tilde{\tau}_{\alpha,\kappa} = 0$. Thus, it follows that the parts $\tilde{\tau}_{\alpha,\kappa}^{-\pm}$ of the hopping $\tilde{\tau}_\alpha$ that originate exclusively from odd field components $\kappa^{-\pm}$ vanish:

$$\tilde{\tau}_{\alpha,\kappa}^{-\pm} = 0.$$

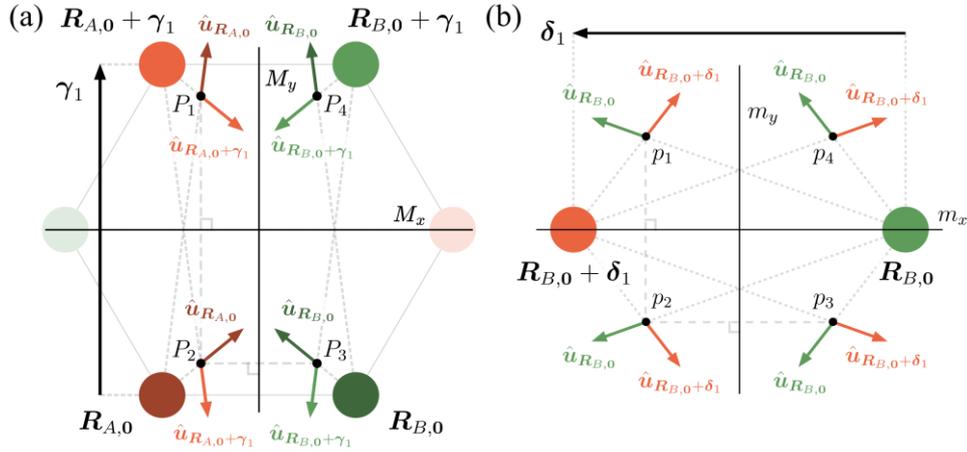

**Fig. 8** – (a)/(b) Unit vectors $\hat{\mathbf{u}}_{\alpha,\mathbf{n}}$ from second/first neighbors in the hexagonal unit cell to the mirror-symmetric points $P_i / p_i$ with $i = 1, \ldots, 4$ [see Eq. (A5)]. The figure illustrates geometrically the unit vector relations expressed in Eqs. (C1), (C3), (C6), and (C7). For example, Eq. (C1) can be visualized by comparing the unit vectors at points $P_1$ and $P_2$ which are related by a reflection about the $M_x$ line. For the magnetic field amplitudes $h_{\alpha,\mathbf{n}}$, the constraints follow from the fact that the distance of two mirror-symmetric points $P$ to the corresponding mirror-symmetry scattering centers is unchanged under a reflection.

Let us now focus our attention on the transformation $\mathbf{r} \to \mathbf{r}' = M_y \cdot \mathbf{r}$, i.e., $(x,y) \to (-x,y)$. We see from Fig. 8(a) that

$$\hat{\mathbf{u}}_{\mathbf{R}_{A,0}}(\mathbf{r}) = M_y \cdot \hat{\mathbf{u}}_{\mathbf{R}_{B,0}}(\mathbf{r}'),$$
$$\hat{\mathbf{u}}_{\mathbf{R}_{A,0}+\gamma_1}(\mathbf{r}) = M_y \cdot \hat{\mathbf{u}}_{\mathbf{R}_{B,0}+\gamma_1}(\mathbf{r}'),$$
$$h_{\mathbf{R}_{A,0}}(\mathbf{r}) = h_{\mathbf{R}_{B,0}}(\mathbf{r}'),$$
$$h_{\mathbf{R}_{A,0}+\gamma_1}(\mathbf{r}) = h_{\mathbf{R}_{B,0}+\gamma_1}(\mathbf{r}').$$
(C3)

Thus, the hopping integral $\tilde{\tau}_A$ can be expressed as:

$$\tilde{\tau}_{A,\kappa} = +\mathcal{N}^{-1} \int_{\mathbb{R}^2} d^2\mathbf{r}' \, \kappa(M_y \cdot \mathbf{r}') \, h_{\mathbf{R}_{B,0}} h_{\mathbf{R}_{B,0}+\gamma_1} \left( \left( M_y \cdot \hat{\mathbf{u}}_{\mathbf{R}_{B,0}} \right) \times \left( M_y \cdot \hat{\mathbf{u}}_{\mathbf{R}_{B,0}+\gamma_1} \right) \right) \cdot \hat{\mathbf{z}}$$
$$= -\mathcal{N}^{-1} \int_{\mathbb{R}^2} d^2\mathbf{r}' \, \kappa(M_y \cdot \mathbf{r}') \, h_{\mathbf{R}_{B,0}} h_{\mathbf{R}_{B,0}+\gamma_1} \left( \hat{\mathbf{u}}_{\mathbf{R}_{B,0}} \times \hat{\mathbf{u}}_{\mathbf{R}_{B,0}+\gamma_1} \right) \cdot \hat{\mathbf{z}}.$$
(C4)

When the gyromagnetic field exhibits odd (even) symmetry with respect to the mirror $M_y$, such that $\kappa(M_y \cdot \mathbf{r}) = -\kappa(\mathbf{r})$ ($\kappa(M_y \cdot \mathbf{r}) = +\kappa(\mathbf{r})$), the above formula implies that $\tilde{\tau}_{A,\kappa} = +\tilde{\tau}_{B,\kappa}$ ($\tilde{\tau}_{A,\kappa} = -\tilde{\tau}_{B,\kappa}$). The same relation can be expressed in terms of hopping components as:

$$\tilde{\tau}_{A,\kappa}^{+\pm} = \mp \tilde{\tau}_{B,\kappa}^{+\pm},$$
(C5)

i.e., the hopping parameters for the two sublattices are additive-symmetric (identical) if the gyromagnetic field is even (odd) under an $M_y$-reflection.

Next, we discuss the nonreciprocal hopping between first neighbors. In particular, we explain why this hopping parameter vanishes for the odd-odd, even-odd, and even-even parity classes. As before, for simplicity, we focus on the gyromagnetic contribution $\tau_\kappa$; $\tau_v$ has similar properties and for conciseness we omit a detailed discussion.

This time, consider the mirror transformations $\mathbf{r} \to \mathbf{r}' = m_x \cdot \mathbf{r}$ and $\mathbf{r} \to \mathbf{r}'' = m_y \cdot \mathbf{r}$, where $m_x$ is the line joining the neighbors and $m_y$ is its perpendicular bisector [see Fig. 8(b)]. The axes $m_{x,y}$ are different from $M_{x,y}$, but are related by symmetry transformations of the modified graphene. Figure 8(b) illustrates that

$$\hat{\mathbf{u}}_{\mathbf{R}_{B,0}}(\mathbf{r}) = m_x \cdot \hat{\mathbf{u}}_{\mathbf{R}_{B,0}}(\mathbf{r}'),$$
$$\hat{\mathbf{u}}_{\mathbf{R}_{B,0}+\delta_1}(\mathbf{r}) = m_x \cdot \hat{\mathbf{u}}_{\mathbf{R}_{B,0}+\delta_1}(\mathbf{r}'),$$
$$h_{\mathbf{R}_{B,0}}(\mathbf{r}) = h_{\mathbf{R}_{B,0}}(\mathbf{r}'),$$
$$h_{\mathbf{R}_{B,0}+\delta_1}(\mathbf{r}) = h_{\mathbf{R}_{B,0}+\delta_1}(\mathbf{r}'),$$
(C6)

and

$$\hat{\mathbf{u}}_{\mathbf{R}_{B,0}}(\mathbf{r}) = m_y \cdot \hat{\mathbf{u}}_{\mathbf{R}_{B,0}+\delta_1}(\mathbf{r}''),$$
$$h_{\mathbf{R}_{B,0}}(\mathbf{r}) = h_{\mathbf{R}_{B,0}+\delta_1}(\mathbf{r}'').$$
(C7)

Using these expressions, we note that the hopping integral satisfies the relations

$$\tau_\kappa = +\mathcal{N}^{-1} \int_{\mathbb{R}^2} d^2\mathbf{r}' \, \kappa(m_x \cdot \mathbf{r}') h_{\mathbf{R}_{B,0}} h_{\mathbf{R}_{B,0}+\delta_1} \left( \left( m_x \cdot \hat{\mathbf{u}}_{\mathbf{R}_{B,0}} \right) \times \left( m_x \cdot \hat{\mathbf{u}}_{\mathbf{R}_{B,0}+\delta_1} \right) \right) \cdot \hat{\mathbf{z}}$$
$$= -\mathcal{N}^{-1} \int_{\mathbb{R}^2} d^2\mathbf{r}' \, \kappa(m_x \cdot \mathbf{r}') h_{\mathbf{R}_{B,0}} h_{\mathbf{R}_{B,0}+\delta_1} \left( \hat{\mathbf{u}}_{\mathbf{R}_{B,0}} \times \hat{\mathbf{u}}_{\mathbf{R}_{B,0}+\delta_1} \right) \cdot \hat{\mathbf{z}},$$
(C8)

and

$$\tau_\kappa = +\mathcal{N}^{-1} \int_{\mathbb{R}^2} d^2\mathbf{r}'' \, \kappa(m_y \cdot \mathbf{r}'') h_{\mathbf{R}_{B,0}+\delta_1} h_{\mathbf{R}_{B,0}} \left( \left( m_y \cdot \hat{\mathbf{u}}_{\mathbf{R}_{B,0}+\delta_1} \right) \times \left( m_y \cdot \hat{\mathbf{u}}_{\mathbf{R}_{B,0}} \right) \right) \cdot \hat{\mathbf{z}}$$
$$= +\mathcal{N}^{-1} \int_{\mathbb{R}^2} d^2\mathbf{r}'' \, \kappa(m_y \cdot \mathbf{r}'') h_{\mathbf{R}_{B,0}} h_{\mathbf{R}_{B,0}+\delta_1} \left( \hat{\mathbf{u}}_{\mathbf{R}_{B,0}} \times \hat{\mathbf{u}}_{\mathbf{R}_{B,0}+\delta_1} \right) \cdot \hat{\mathbf{z}}.$$
(C9)

Hence, we conclude that $\tau_\kappa$ vanishes when the gyromagnetic field has even $m_x$-parity or odd $m_y$-parity. If the $M_{x,y}$-parities are well-defined, those with respect to the axes $m_{x,y}$ are well-defined too. This is portrayed in Fig. 5 of the main text, as evidenced by the highlighted reflection symmetries and antisymmetries of the unit cells. In fact, the 3-fold rotational symmetry about the center of the cell determines that the $m_x(m_y)$ and $M_x(M_y)$ parities are identical. Thus, it follows that $\tau_\kappa$ vanishes when the field $\kappa(\mathbf{r})$ has even $M_x$-parity or odd $M_y$-parity. This means that this hopping can be nontrivial only for the odd-even symmetry class.

## Appendix D: Gap Chern Number

The nonreciprocally perturbed graphene has a photonic bandgap when the split term $m$ of the Hamiltonian $\mathcal{H}'_\mathbf{k}$ is nonvanishing. The gap Chern number is given by

$$\mathcal{C}_{\text{gap}} = \frac{1}{2\pi} \int_{\text{FBZ}} d\mathbf{k} \left( \nabla_{\mathbf{k}} \times \mathbf{A}_{\mathbf{k}} \right) \cdot \hat{\mathbf{z}}, \tag{D1}$$

where $\mathbf{A}_{\mathbf{k}} = \text{Re}\, i \langle \varphi_{\mathbf{k}} | \nabla_{\mathbf{k}} \varphi_{\mathbf{k}} \rangle / \langle \varphi_{\mathbf{k}} | \varphi_{\mathbf{k}} \rangle$ is the Berry potential expressed in terms of the unnormalized ground state $\varphi_{\mathbf{k}}$ of $\mathcal{H}'_{\mathbf{k}}$. We denote the gradient in quasimomentum space by $\nabla_{\mathbf{k}}$ and use the canonical inner-product in $\mathbb{C}^2$. From Eqs. (6), (7), (17), (18), and (19), we can write the tight-binding Hamiltonian of the modified graphene explicitly as

$$\mathcal{H}'_{\mathbf{k}} = \left[ 1 - \Omega - \tilde{t} g(\mathbf{k}) - s h(\mathbf{k}) \right] \mathbf{1}_{2\times 2} + \begin{pmatrix} -m h(\mathbf{k}) & -T^* f^*(\mathbf{k}) \\ -T f(\mathbf{k}) & m h(\mathbf{k}) \end{pmatrix}. \tag{D2}$$

The functions $f(\mathbf{k})$, $g(\mathbf{k})$, and $h(\mathbf{k})$ are defined in the main text [see Eqs. (8) and (20)] and $T = t + i\tau$. The unnormalized eigenvector for the lowest frequency band may be chosen as

$$\varphi_{\mathbf{k}} = \begin{pmatrix} T^* f^*(\mathbf{k}) \\ -m h(\mathbf{k}) + \sqrt{|T|^2 |f(\mathbf{k})|^2 + m^2 h^2(\mathbf{k})} \end{pmatrix}. \tag{D3}$$

Evidently, this gauge is smooth at any point of the FBZ except possibly at the zeros of the function $f(\mathbf{k})$, which are the Dirac points $\text{K}^{\pm}$ [Eq. (10)]. Indeed, in such a case one can have $\varphi_{\mathbf{k}} = 0$ depending on the sign of $m h(\mathbf{k})$. Thereby, the Berry potential is also smooth on the set $\text{FBZ} \setminus \{\text{K}^+, \text{K}^-\}$. This means that we can use Stokes theorem to rewrite the gap Chern number in Eq. (D1) as

$$\mathcal{C}_{\text{gap}} = -\frac{1}{2\pi} \lim_{q \to 0} \left( \int_0^{2\pi} d\phi\, q\, \mathbf{A}_{\text{K}^+ + \mathbf{q}} \cdot \hat{\boldsymbol{\phi}} + \int_0^{2\pi} d\phi\, q\, \mathbf{A}_{\text{K}^- + \mathbf{q}} \cdot \hat{\boldsymbol{\phi}} \right), \tag{D4}$$

i.e., in terms of line integrals of the Berry potential over circumferences concentric with the Dirac points and with infinitesimal radii.

Next, we consider a system of polar coordinates $\mathbf{q} = (q, \phi)$ centered at the Dirac points. Near $\text{K}^{\pm}$, the ground state in Eq. (D3) can be approximated by

$$\boldsymbol{\varphi}_{K^{\pm}+\mathbf{q}} \approx \begin{pmatrix} T^{*}\nabla_{\mathbf{k}}f^{*}\left(K^{\pm}\right)\cdot\mathbf{q} \\ -mh\left(K^{\pm}\right)\left(1\pm\operatorname{sgn}(m)\right) \end{pmatrix}, \tag{D5}$$

to leading-order in $\mathbf{q}$. In writing the second component of the vector, we take into account that $h\left(K^{\pm}\right) = \mp 3\sqrt{3}$. Using $\nabla_{\mathbf{k}}f^{*}(K^{\pm}) = 3(i,\pm 1)^{\mathrm{T}}/2$, we obtain the corresponding Berry potential

$$\mathbf{A}_{K^{\pm}+\mathbf{q}} = \pm\frac{\alpha q}{\alpha q^{2} + \beta\left(1\pm\operatorname{sgn}(m)\right)}\hat{\boldsymbol{\phi}}, \tag{D6}$$

where $\alpha = 9|T|^{2}/4$ and $\beta = 54m^{2}$. Feeding this result to Eq. (D4) yields $\mathcal{C}_{\text{gap}} = -\left(I^{+} + I^{-}\right)$ with

$$I^{\pm} = \pm\lim_{q\to 0}\frac{\alpha q^{2}}{\alpha q^{2} + \beta\left(1\pm\operatorname{sgn}(m)\right)}. \tag{D7}$$

Noting that $I^{+} = +1$ and $I^{-} = 0$ when $m < 0$, and $I^{+} = 0$ and $I^{-} = -1$ when $m > 0$, we conclude that $\mathcal{C}_{\text{gap}} = \operatorname{sgn}(m)$, in agreement with Eq. (28) of the main text.